\documentclass[prd,aps,twocolumn,a4paper,showkeys,nofootinbib]{revtex4-1}

\usepackage{graphicx,psfrag}
\usepackage{mathrsfs}
\usepackage{amsmath,amsfonts,amssymb}
\usepackage{multirow}
\usepackage{comment}
\usepackage{ulem}
\usepackage{multirow}
\usepackage{hyperref}

\newcommand{\be}{\begin{equation}}
\newcommand{\ee}{\end{equation}}
\newcommand{\bea}{\begin{eqnarray}}
\newcommand{\eea}{\end{eqnarray}}
\newcommand{\bel}{\begin{align}}
\newcommand{\eel}{\end{align}}

\def\data{{\text{\textbf d}}}
\def\signal{{\text{\textbf s}}}
\def\noise{{\text{\textbf n}}}

\def\half{\frac{1}{2}}
\def\e{{\rm e}}

\def\d{{\rm d}}

\def\B{\mathcal{B}}

\def\Msun{{\rm M_{\odot}}}

\def\freeparams{{\params_{\rm free}}}
\def\extparams{{\params_{\rm ext}}}
\def\kt{\kappa^{\rm T}_2}

\def\params{\boldsymbol{\theta}}
\def\recalib{\boldsymbol{\delta}}
\def\GMc2{{\rm G M_{\odot} c^{-2}}}

\def\model{{\tt NRPMw}}
\def\oldmodel{{\tt NRPM}}

\def\spin{{\boldsymbol \chi}}
\def\chieff{{\chi_{\rm eff}}}
\def\adrift{\alpha_{\rm peak}}
\def\recalibpm{{\recalib_{\rm fit}}}
\def\tc{t_{\rm coll}}

\usepackage{pifont} 

\newcommand{\bajes}{{\scshape bajes}}

\usepackage{color}
\definecolor{cyan}{rgb}{0,0.9,0.9}
\definecolor{orange}{rgb}{0.9,0.5,0}
\definecolor{magenta}{rgb}{1,0,1}
\definecolor{purple}{rgb}{0.8,0.4,0.8}
\definecolor{gray}{rgb}{0.8242,0.8242,0.8242}

\newcommand{\paperI}{Paper~{I}}

\begin{document}

\title{Kilohertz Gravitational Waves from Binary Neutron Star Mergers:\\
  Inference of Postmerger Signals with the Einstein Telescope} 

\author{Matteo \surname{Breschi}} \author{Rossella \surname{Gamba}} \author{Ssohrab \surname{Borhanian}} \author{Gregorio \surname{Carullo}} \author{Sebastiano \surname{Bernuzzi}} \affiliation{Theoretisch-Physikalisches Institut,
  Friedrich-Schiller-Universit{\"a}t Jena, 07743, Jena, Germany}

\date{\today}

\begin{abstract}
   Next-generation detectors are expected 
   to be sensitive to postmerger signals 
   from binary neutron star coalescences and thus
   to directly probe the remnant dynamics.
   We investigate the scientific potential of postmerger detections with the Einstein Telescope using 
     full Bayesian analyses with the state of the art waveform model
     {\model}. We find that:
  (i) Postmerger signals with SNR ${\sim}7$ can be confidently detected
  with a Bayes' factor of $\log\B\simeq5$ ($\e$-folded) and
  the posterior distributions 
  report informative
  measurements already at SNR ${\sim}6$
  for some noise realizations.
  (ii) The postmerger peak frequency $f_2$ can be confidently identified at SNR $7$ with errors of 
  $O(1~{\rm kHz})$, that decrease below $O(100~{\rm Hz})$ for SNR 10.
  (iii) The remnant's time of collapse to black hole
  can be constrained to $O(20~{\rm ms})$ at SNR 10.
  However, the inference can be biased by 
  noise fluctuations
  if the latter exceed the signal's amplitude 
  before collapse.
  (iv) Violations of the EOS-insentive relations for $f_2$ can be detected at
  SNR $\gtrsim 8$ if the frequency shifts are 
  $\gtrsim 500~{\rm Hz}$; they can be smoking
  guns for EOS softening effects at extreme densities.
  However,
  the $f_2$ measurement can be 
  significantly biased
  by subdominant frequency components
  for short-lived remnants.
  In these cases, an EOS softening might be better inferred from the remnant's
  earlier collapse.
\end{abstract}


\maketitle

\section{Introduction}
\label{sec:intro}

Third-generation gravitational-wave (GW) observatories, such as Einstein Telescope (ET)~\cite{Hild:2008ng}, will observe gravitational radiation from binary neutron star (BNS) remnants. The detection of postmerger (PM) gravitational transients will give the most direct access to the merger outcome by probing the nature of the compact object in the remnant, see e.g. \cite{Radice:2020ddv,Bernuzzi:2020tgt} for recent reviews. 
Moreover, 
if the remnant does not promptly collapse into black hole (BH),
the PM radiation can provide information on the extreme density equation of state (EOS) of the neutron star (NS) remnant,
thus impacting the nuclear physics of NSs~\cite{Chatziioannou:2017ixj,Tsang:2019esi,Breschi:2019srl,Easter:2020ifj,Breschi:2021xrx,Wijngaarden:2022sah}.
In Ref.~\cite{Breschi:2022} ({\paperI} hereafter) we have developed a
novel PM model built on a wavelet basis and incorporating several
NR-driven quasiuniversal relations. Here, we expand our investigations
by exploring the capabilities of ET in Bayesian parameter estimation
of PM signals. A follow-up paper of this series will report on full-spectrum analysis.

The Einstein Telescope~\cite{Hild:2008ng,Hild:2010id,
Hild:2011np,Punturo:2010zz,Punturo:2010zz,Punturo:2010zza,Maggiore:2019uih,
Sathyaprakash:2011bh,Sathyaprakash:2012jk,Amann:2020jgo} 
is a triangularly-shaped underground infrastructure for GW observations,
with a photodetector at each vertex measuring GW fluctuations with interferometric techniques and
employing laser beams running along the adjacent sides at a 60 degree angle.
Hence, the observatory is composed by three detectors.
ET will improve the sensitivities of modern observatories
by increasing the size of the interferometer from the $3~{\rm km}$ arm length 
of the Virgo detector 
to $10~{\rm km}$, and by implementing a series of new technologies. These include a cryogenic system, 
quantum technologies to reduce the laser fluctuations
and noise-mitigation measures to reduce environmental perturbations.
Moreover, the instrument will exploit a {\it xylophone-design}
in which each GW detector is composed of two individual interferometers:
a cryogenic low-power interferometer with sensitivity plateau 
in the low-frequency band,
roughly between $7~{\rm Hz}$ and $20~{\rm Hz}$, 
and an high-power interferometer that covers the higher portion 
of the spectrum, roughly above ${\sim}100~{\rm Hz}$.
The ET-D configuration discussed in~\cite{Hild:2010id} 
and employed in this work has a sensitivity bucket covering the GW spectrum from ${\sim}5~{\rm Hz}$ up to ${\sim }3~{\rm kHz}$.
Such broad-band sensitivity is ideal for BNS observations.
Low-frequency measurements follow the binary evolution for many cycles before merger
extracting precise measurements of the progenitor's properties.
In contrast, the high-frequency end, i.e. $\gtrsim 1~{\rm kHz}$, 
enables the detection of PM signals at sensitivities that are
unreachable by current infrastructures. 

The significant complexity of the PM source emission implies that 
standard PE techniques, relying on extremely accurate models
parametrised by the minimal set of the system's degrees of freedom,
can no longer be applied.
For this reason, recent Bayesian studies of BNS PM transients 
employed largely different modeling choices, ranging from signal-agnostic reconstructions
to semi-agnostic descriptions calibrated on numerical simulations.data
The advantage of agnostic techniques lies in an unbiased reconstruction of the signal,
coming however at the price of losing information content buried in the noise.
Semi-agnostic reconstructions, which can directly incorporate information
from NR simulations, allow to dig deeper into the detector background and
hence detect signals at lower signal-to-noise ratio (SNR). 
However such measurements require extensive 
training datasets and the inclusion of all the relevant contributions in order
to provide a faithful recovery of the signal's properties.
An additional complication that arises when comparing different semi-agnostic approaches
thus stems in the NR dataset used to built and validate them.

\citet{Chatziioannou:2017ixj} performed
PE studies of BNS PM transients
with model-independent Bayesian techniques,
showing informative posteriors 
for PM SNR
of ${\sim}5$ with a validation set of three NR simulations.
However, the method employed in 
Ref.~\cite{Chatziioannou:2017ixj} did not 
provide estimation of the evidence,
making difficult a more direct 
and complete comparison
with different techniques.
\citet{Tsang:2019esi} used a simplified model based 
on damped sinusoidal template that aims to characterize the 
dominant PM peak.
This method recovered informative measurements for a 
threshold of SNR between $4$
and $8$ using a validation set of four binaries.
\citet{Breschi:2019srl} constructed the PM model
employing NR-calibrated relations
that naturally permits the attachment to pre-merger templates,
introducing the first 
inspiral-merger-postmerger BNS model.
This approach show detectabillity 
threshold for PM SNR ${\sim}8$
over a NR set of 10 binaries.
The performances are improved in Ref.~\cite{Breschi:2021xrx} to SNR $7$,
corresponding to a Bayes' factor (BF)
of $\e^{5}$,
including model recalibration parameters in the PE
routine
(see also {\paperI}).
Similarly to \cite{Tsang:2019esi},
\citet{Easter:2020ifj} modeled the PM spectrum employing a
superposition of multiple damped sinusoidal components
recovering PM transients for SNR 
$7$ over a validation set of nine binaries,
which corresponds to a BF of $\e^{8}$.
\citet{Wijngaarden:2022sah} extended the approach of
Ref.~\cite{Chatziioannou:2017ixj} by including the pre-merger portion
of data. 

In this paper, 
we employ the novel {\model} model~\cite{Breschi:2022}
for Bayesian studies of GW signals from BNS PM remnants
with the next-generation detector ET,
investigating the PE performances
under different noise realizations.
The paper is structured as follows.
In Sec.~\ref{sec:framework}, we 
introduce the framework
employed for producing our set of simulated signals, and our PE configurations.
We perform a consistent simulation-recovery
study with {\model} in Sec.~\ref{sec:template}.
We discuss PE studies 
on NR data in Sec.~\ref{sec:nr},
inspecting the detectability performances 
and the recovered posteriors on the signal parameters.
In Sec.~\ref{sec:softness},
we reanalyze the softening case presented in \cite{Radice:2016rys,Breschi:2019srl} investigating 
extreme-matter tests for non-hadronic EOS.
We conclude in Sec.~\ref{sec:conclusion}.

\paragraph*{Conventions --}

All quantities are expressed in SI  units,
with masses in Solar masses $\Msun$
and distances in Mpc.
Henceforth, the symbol `$\log$' denotes the natural logarithm.
The total binary mass is indicated with $M= m_1 +
m_2$, the mass ratio $q = m_1 /m_2 \ge 1$, and the 
symmetric mass ratio $\nu = m_1 m_2 / M^2$.
The dimensionless spin vectors are denoted with $\spin_i$ for $i=1,2$
and
the spin component 
aligned with the orbital angular momentum $\textbf{L}$
are labeled as $\chi_i = \spin_{i}\cdot \textbf{L} / |\textbf{L}|$.
The effective spin parameter $\chieff$ is
a mass-weighted, aligned spin combination, i.e.
\be
\label{eq:chieff}
\chieff = \frac{m_1 \chi_{1}+m_2 \chi_{2}}{M}\,.
\ee
Moreover,
the quadrupolar tidal deformability parameters are defined 
as $\Lambda_{i}=({2}/{3})\,k_{2,i}\,C_i^{-5}$ for $i=1,2$,
where $k_{2,i}$ and $C_i$ are the second Love number and the 
compactness of the $i$-th object, respectively.
The quadrupolar tidal polarizability $\kt$ is
\be
\label{eq:k2t}
\kt =3\nu\,\left[\left(\frac{m_1}{M}\right)^3 \Lambda_1 + (1\leftrightarrow 2)\right]\,.
\ee
Masses, spins, and tides define the 
intrinsic parameters of a BNS system, 
i.e. $\params_{\rm bin}=\{M,q,\chi_{1},\chi_{2},\Lambda_1,\Lambda_2\}$.
The extrinsic parameters of the source $\extparams=\{D_L, \iota , \alpha, \delta, \psi, t_{\rm mrg},\phi_{\rm mrg}\}$,
i.e. luminosity distance $D_L$,
inclination angle $\iota$,
right ascension angle $\alpha$,
declination angle $\delta$,
polarization angle $\psi$,
time of coalescence $t_{\rm mrg}$, and phase at the merger $\phi_{\rm mrg}$,
allow us to identify the location and orientation of the source.
The PM model {\model} is parameterized by 
two additional sets of degrees of freedom
(see {\paperI} for a detailed discussion).
The PM parameters 
$\freeparams=\{\phi_{\rm PM},\tc,\adrift\}$
correspond to:
the PM phase-shift $\phi_{\rm PM}$ that 
identifies the phase discontinuity after merger;
the time of collapse $\tc$ that characterizes
the collapse of the remnant into BH; and 
the frequency drift $\adrift$
that accounts for linear evolution
of the dominant $f_2$ component.
Moreover, we include
the recalibration parameters $\recalibpm$
that account for deviations from the
predictions of the EOS-insensitive 
relations consistently with the 
related theoretical uncertainties.

\section{Framework}
\label{sec:framework}

In this section,
we discuss the framework 
employed to generate and analyze the artificial data, called ``injections''.
We discuss the injections creation in Sec.~\ref{sec:inject},
the PE settings are reported in Sec.~\ref{sec:bayes}
and we discuss 
the treatment of different noise realizations
in Sec.~\ref{sec:noiserealiz}.
All the analyses are performed with the publicly available
  \bajes{} pipeline~\cite{Breschi:2021wzr}.

\subsection{Injection settings}
\label{sec:inject}

We generate artificial data 
for the triangular, triple-interferometer ET detector~\cite{Hild:2008ng,Punturo:2010zz,Maggiore:2019uih}, 
segmented into chunks of $1~{\rm s}$ duration with a sampling rate of $16384~{\rm Hz}$.
For each detector $i$, 
the artificial data series
$d_i(t)=s_i(t)+n_i(t)$ 
is composed of the signal $s_i(t)$ projected onto the $i$-th detector
and the respective noise contribution $n_i(t)$.
We label with bold symbols the sets
$\signal(t)=\{s_i(t)\}$,
$\noise(t)=\{n_i(t)\}$, and $\data(t)=\{d_i(t)\}$
such that $\data(t)=\signal(t)+\noise(t)$.
The injected signals vary depending on the 
study:
in Sec.~\ref{sec:template},
the signal is taken to be identical to {\model}
for a fixed set of parameters
in order to perform a consistent injection-recovery test;
while, in Sec.~\ref{sec:nr},
we inject NR templates in order to study the performance
of {\model} in a more realistic scenario.
The noise is assumed to be Gaussian, wide-sense stationary,
and colored according to the power spectral density (PSD) 
expected for ET-D~\cite{Hild:2011np}.

Given a data series $d$ 
and a template $h$, 
we introduce the matched-filtered
SNR $\rho$ as
\be
\label{eq:snr}
\rho(h) = \frac{(d|h)}{\sqrt{(h|h)}}\,,
\ee
where the inner product $(\cdot|\cdot)$
between two time-series,
say $a(t)$ and $b(t)$,
corresponds to
\be
\label{eq:innerprod}
(a|b) = 4 \Re \int \frac{a^*(f) \,b(f)}{S_n(f)}\, \d f\,,
\ee
where 
$a(f)$ and $b(f)$ are the Fourier transforms respectively of $a(t)$ and $b(t)$,
and $S_n(f)$ is the PSD employed to generate the noise segments.
The definition Eq.~\eqref{eq:snr} can be extended to multiple detectors 
employing quadrature summation.
Then, we label $\rho_{\rm inj} = \rho(s)$
the ratio computed with the exact injected template; while,
$\rho_{\rm rec} = \rho(h_\text{\model})$
is the SNR recovered with the {\model} template $h_\text{\model}$.
The templates $s(t)$ are injected
at seven different SNRs 
$\rho_{\rm inj}=\{5, 6, 6.5, 7, 7.5, 8, 10\}$,
which corresponds to
locating the binaries at different luminosity distances.
Moreover, the simulated binaries are oriented with  
$\iota=0$, $\psi=0$, and optimal sky position
for the employed detector
$\{\alpha = 2.621, \delta = 0.706\}$.
Moreover, each artificial signal is analyzed 
employing five random noise realizations 
(see Sec.~\ref{sec:noiserealiz}).

\subsection{Parameter estimation}
\label{sec:bayes}

The PE studies on the artificial data are performed 
with the nested sampling algorithm {\scshape ultranest}~\cite{Buchner2021}
with {3000} initial live points.
For a fixed set of data $\data(t) $
the likelihood function $p(\data|\params,H_{\rm S})$~\cite{Veitch:2014wba,Breschi:2021wzr}
is given by
\be
\label{eq:likelihood}
\log p(\data|\params,H_{\rm S}) \propto -\half \sum_i \big(d-h(\params)\big|d-h(\params)\big)_i\,,
\ee
where $H_{\rm S}$ represents the signal hypothesis.
The subscript $i$ runs over the employed detectors
and it denotes that the inner products are estimated 
with the corresponding data series, projected waveform, 
and PSD for each detector.
The inner product in Eq.~\eqref{eq:likelihood} is integrated 
over the frequency range $[1,8]~{\rm kHz}$, 
in order to isolate the contribution from the PM signal.
We perform analytical marginalization
over reference time $t_{\rm mrg}$ and phase $\phi_{\rm mrg}$.
Then, the information on the parameters $\params$ is encoded 
in the posterior distribution $p(\params|\data,H_{\rm S}) $, that can be estimated
with Bayes' theorem as
\be
\label{eq:posterior}
p(\params|\data,H_{\rm S}) \propto p(\data|\params,H_{\rm S}) \,p(\params|H_{\rm S}) \,.
\ee
Resorting to a nested sampling algorithm~\cite{Skilling:2006},
allows to straightforwardly and accurately estimate the evidence integral:
\be
\label{eq:evidence}
p(\data|H_{\rm S})= \int p(\data|\params,H_{\rm S}) \,p(\params|H_{\rm S}) \,\d\params \,,
\ee
that in turns allows to compute the BF ($\B$) 
of the signal hypothesis against the noise hypothesis as 
$\B = p(\data|H_{\rm S})/p(\data|H_{\rm N})$.
The noise hypothesis $H_{\rm N}$ corresponds to the assumption 
that no signal is included in the recorded data.
We fix the {\it nominal} detectability threshold at $\log\B=5$~\cite{Kass:1995}.

The sampling is performed in  
total mass $M\in[1, 6]~\Msun$ and mass ratio $q\in [1,2]$, 
differently to what is presented in~\cite{Breschi:2019srl,Breschi:2021wzr},
due to the analytical form of the empirical relations.
In order to maintain a uniform prior in the mass components $m_{1,2}$,
the prior $p(M,q|H_{\rm S})$ is modified according to~\cite{Callister:2021gxf}.
We include aligned spin parameters and employ
an isotropic prior with the constraint 
$|\chi_{i}|\le 0.2$ for $i=1,2$.
Tidal parameters $\Lambda_1$ and $\Lambda_2$
are sampled with a uniform prior in the range $[0,4000]$.
For the luminosity distance $D_L$, we use a volumetric prior
in the range $[20, 500]~{\rm Mpc}$
in order to confidently include the 
injected values.
The remaining extrinsic parameters are treated according to~\cite{Breschi:2021wzr}.
Moreover,
we include the PM parameters $\freeparams$ in the PE routine and
perform the sampling in the mass-scaled quantities
using a uniform prior distribution for
$\tc/M\in[t_0/M, 3000]$,
$M^2\adrift\in[-10^{-5}, 10^{-5}]$ and 
$\phi_{\rm PM}\in[0, 2\pi]$.
Finally,
we introduce in the sampling
the recalibration parameters $\recalibpm$
in order to account the intrinsic errors
of the calibrated formulae.
For these terms, we employ a normally distributed 
prior with zero mean and variance defined by the 
estimated relative errors~(see {\paperI}).

\subsection{Noise realizations}
\label{sec:noiserealiz}

The near-threshold SNR of the signals under consideration requires additional care
when extracting information from a simulation study.
As such, in order to investigate the impact of noise fluctuations,
we generate artificial data $\data_k(t)=\signal(t)+\noise_k(t)$ by
injecting the targeted template $\signal(t)$ 
into different random noise realizations $\noise_k=\{n_i(t)\}_k$,
where 
$i$ runs over the employed detectors and
$k$ runs over the noise realizations.
We employ a total of five different noise realizations
fixing the initialization seed of the pseudo-random number generator 
in the {\scshape bajes} pipeline
\footnote{In particular, we employ the following 
				random seeds for the 
				{\tt bajes\_inject} routine: $\{38, 170817, 742435, 822959, 4420301 \}$.}.
The PE studies are performed on all the included realizations $k$ for each signal $s(t)$ and every injected SNR.

Once the posterior distributions 
$p(\params|\data_k,H_{\rm S})$ are estimated
for each $k$,
we compute an overall posterior 
in order to average over the different noise realizations.
The overall posterior distribution is computed 
equally weighting each noise realization 
and averaging the recovered posteriors,
i.e.
\be
\label{eq:equal-weighted-prob}
p(\params|\signal, H_{\rm S}) \propto \sum_{k} p(\params|\data_k,H_{\rm S})\,,
\ee
where $k$ runs over the employed noise realizations.
This approach aims to estimate an agnostic
and comprehensive posterior distribution
that correctly incorporates the full statistical uncertainties.
As we will show below, such uncertainties are relevant when discussing some of the cases under consideration.
This estimate will improve with a larger number of noise configurations, $k \gg 1$,
and is limited only by the computational cost.

\section{{\model} injections}
\label{sec:template}

 \begin{figure}[t]
	\centering 
	\includegraphics[width=0.49\textwidth]{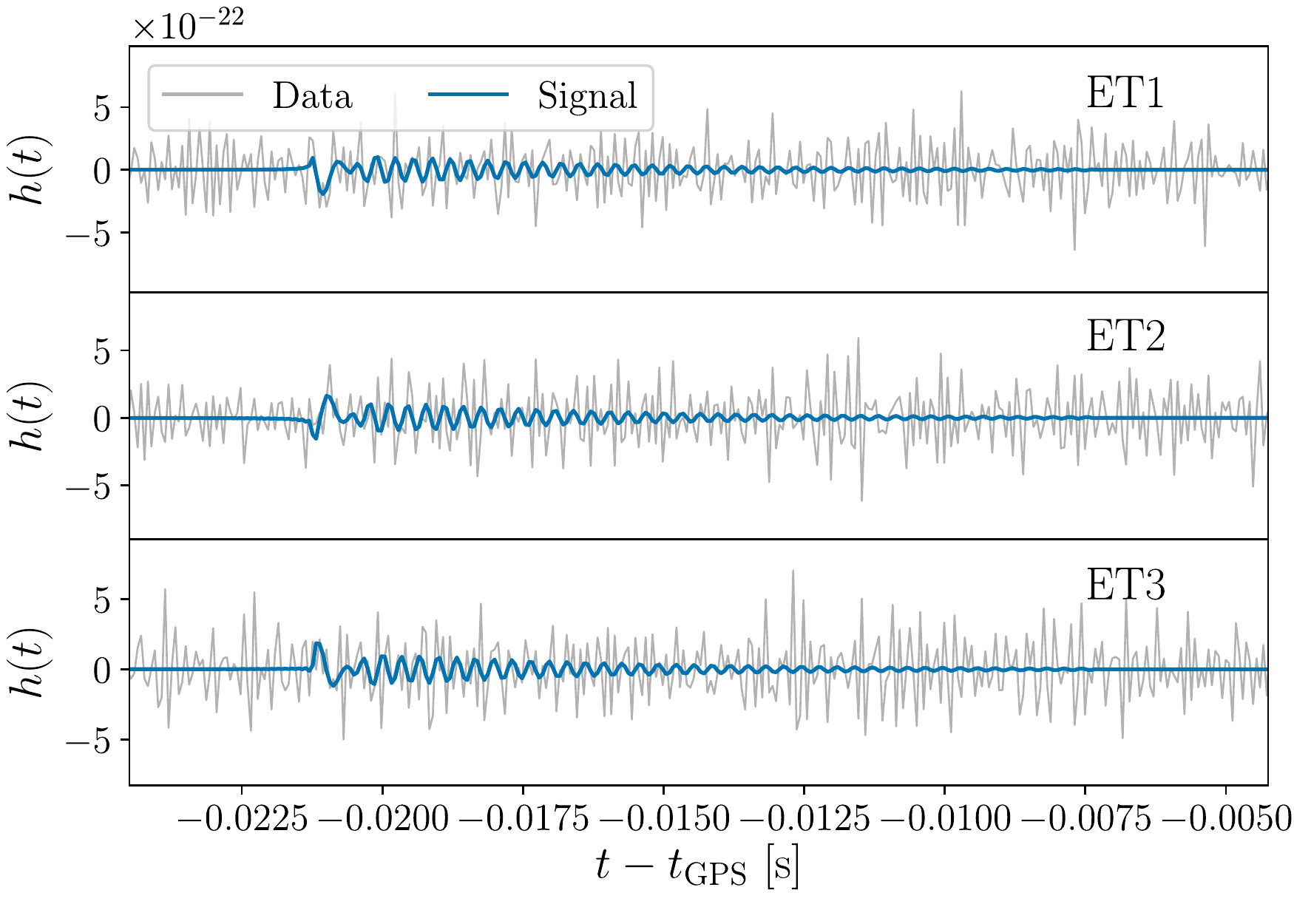}
	\caption{Injected PM signal (blue) and 
					artificial data (gray)
					 for an exemplary noise realization
					 at PM SNR 8.
					 The different panels show 
					 the time series for 
					 three ET interferometers.
					 The signal corresponds to 
					 the {\model} template 
					 computed for the parameters 
					 $M=2.82~\Msun$, 
					 $q=1.5$, 
					 $\kt=90$,
					 $D_L=70~{\rm Mpc}$,
					 $\tc=14~{\rm ms}$,  
					 $\adrift=0.021~{\rm kHz}^2$, 
					 $\phi_{\rm PM}=\pi/2$,
					 and all recalibrations 
					 are identically zero, i.e. $\recalibpm=0$.
				 	The noise segments are generated 
			 		assuming Gaussian and stationary noise 
		 			according to ET-D PSD~\cite{Hild:2011np}.}
	\label{fig:nrpmw-wave}
\end{figure}

A first crucial step necessary to verify the reliability of a model 
and of the underlying Bayesian framework
is a consistent injection-recovery study
in which the simulated signal corresponds to a realization 
(for a given combination of $\params$)
of the same model employed for recovery.
In this way, it is possible to directly investigate 
sampling errors or possible complications due to poor modeling choices.
Consequently, we employ {\model} to generate the ``real'' signal $s(t)$ to be GW170817-like,
setting total mass $M=2.82~\Msun$, mass ratio $q=1.5$, 
tidal polarizability $\kt=90$,
time of collapse $\tc=14~{\rm ms}$, frequency drift $\adrift=0.021~{\rm kHz}^2$, PM phase $\phi_{\rm PM}=\pi/2$,
and all recalibrations identically to zero, $\recalibpm=0$.
Figure~\ref{fig:nrpmw-wave} shows 
the GW signals and the artificial data 
for each ET interferometer 
corresponding to an injection with PM  
SNR 8 for an exemplary noise realization.

We perform two sets of PE studies for this case.
In the first set, 
we do not take into account the recalibrations
$\recalibpm$; 
in the second set, we use recalibration
in the PE analysis.
The comparison between these two sets illustrates
the effect of the recalibrations $\recalibpm$ on the 
recovered posterior distributions
in the case where such recalibration 
is not required to correctly characterize the true signal,
 aiding the interpretation
of the more complicated 
and realistic cases explored in the next sections.
In general, 
it is expected that the inclusion of recalibrations 
will not introduce deviations in the mean values 
but will lead to a widening of the recovered posteriors
due to the fluctuations associated with the calibrated relations, yielding
more conservative measurements.
In Sec.~\ref{sec:nrpmw-detect},
we discuss the recovered BFs and 
detectability threshold;
while, the recovered posterior 
distributions are presented in
Sec.~\ref{sec:nrpmw-pe},
comparing non-recalibrated 
and recalibrated PEs.

\subsection{Detectability}
\label{sec:nrpmw-detect}

 \begin{figure}[t]
	\centering 
	\includegraphics[width=0.49\textwidth]{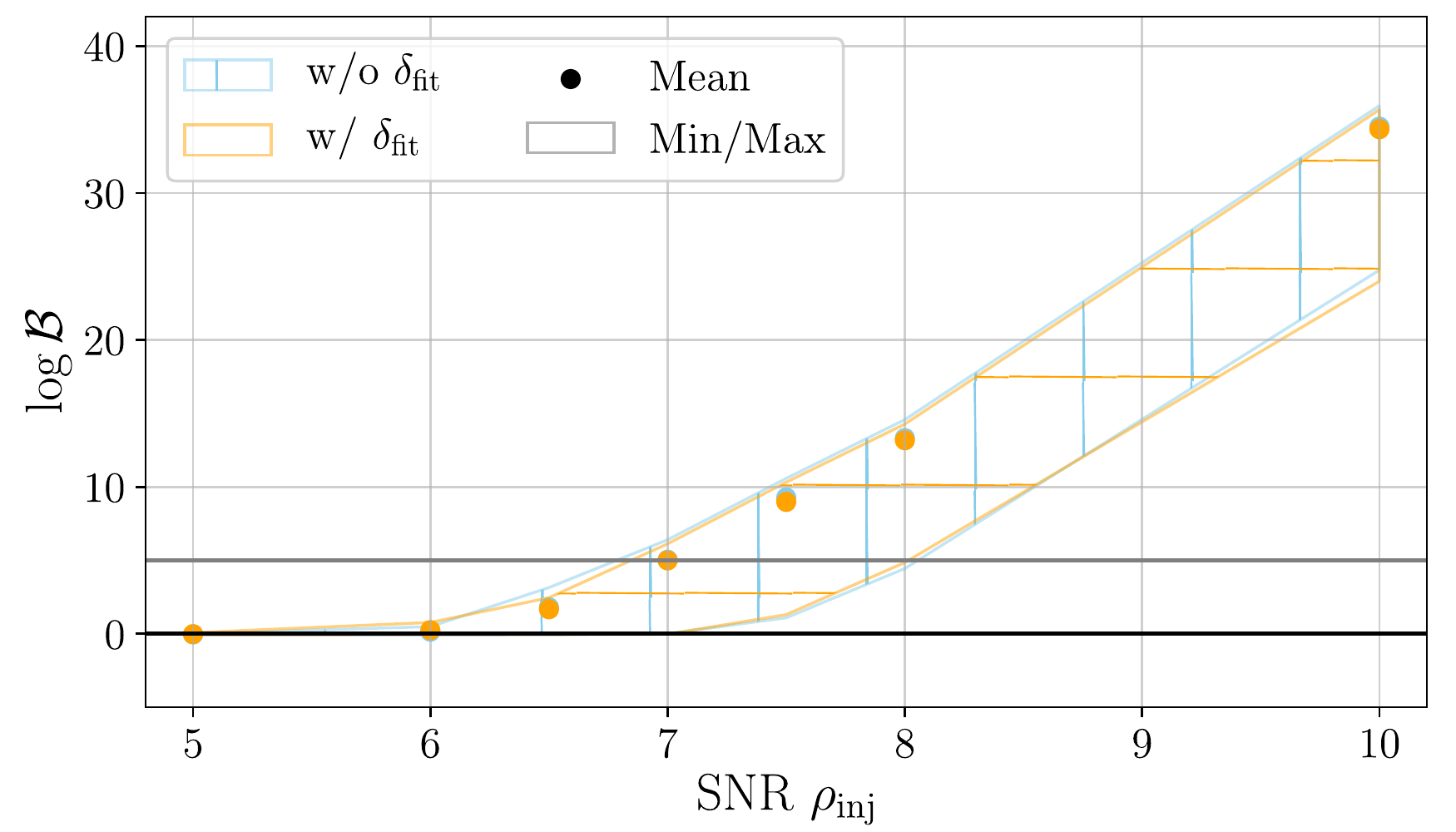}
	\caption{Logarithmic BFs, $\log\B$, as functions of the PM SNR 
		$\rho_{\rm inj}$ of the injected {\model}
		template.
		The analysis without recalibrations $\recalib_{\rm fit}$ is reported in blue; while, the recalibrated 
		study is shown in orange.
		The dots refer to the mean values
		averaged over the different noise realizations
		and the shadowed areas correspond to the 
		minimum and maximum values.
		Two horizontal lines identify $\log\B=0$ (black) and 
		$\log\B=5$ (gray).}
	\label{fig:nrpmw-bfs}
\end{figure}

We investigate the BFs in favor of the signal hypothesis
as function of the PM SNR of the injected transients,
in order to understand the threshold required by {\model} 
to perform an informative inference.

Figure~\ref{fig:nrpmw-bfs} shows 
the recovered BFs 
for the injected binarys as function of the injected
SNR,
where the error-bars are estimated 
using the values recovered from different
noise realization.
In particular, 
we report the median values 
and the uncertainties are computed 
from the maximum and minimum 
estimates.
The overall trend shows non-vanishing $\log\B$
for $\rho_{\rm inj}>6$ and the nominal threshold is
reached at $\rho_{\rm inj}=7$.
However, noise fluctuations can affect the 
recovered BFs, shifting the detectability threshold 
to SNRs of $\rho_{\rm inj}\simeq8$.
Above $\rho_{\rm inj}\simeq8$,
the errors of the recovered log-BFs appears to be 
roughly constant with values of ${\sim}15$
(considering the difference between the minimum and 
the maximum).
Moreover, the recovered BFs are not affected 
by the introduction of the recalibration 
$\recalib_{\rm fit}$, 
since non-recalibrated {\model}, i.e.
without the inclusion of $\recalib_{\rm fit}$,
is capable to fully reproduce the 
injected signal.
The use of recalibrations is expected 
to be more important in the inference of signals that are not perfectly
matched by the template model.

\subsection{Posterior distributions}
\label{sec:nrpmw-pe}

\begin{figure*}[t]
	\centering 
	\includegraphics[width=0.49\textwidth]{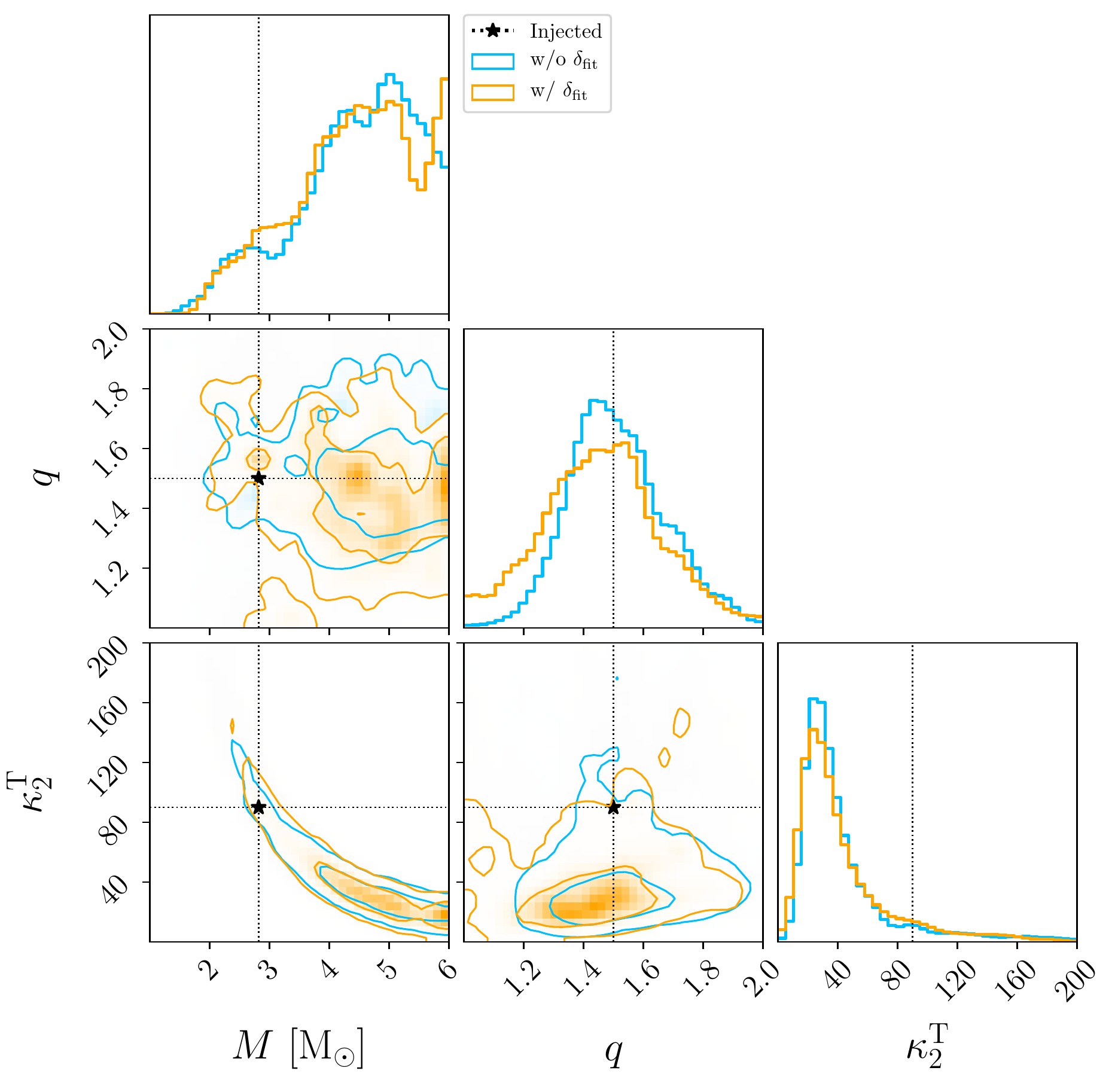}
	\includegraphics[width=0.49\textwidth]{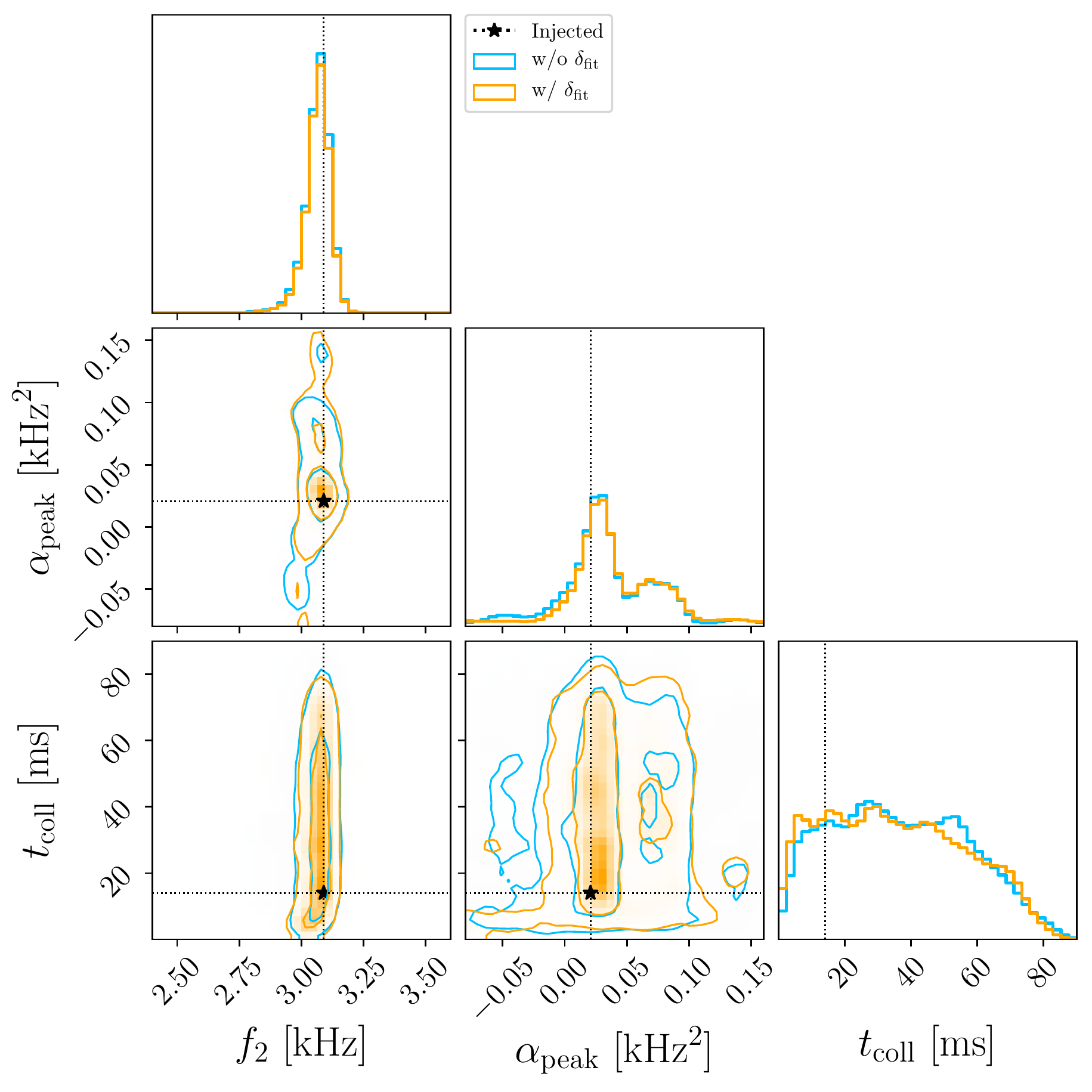}\\
	\vspace{0.4cm}
		\includegraphics[width=0.49\textwidth]{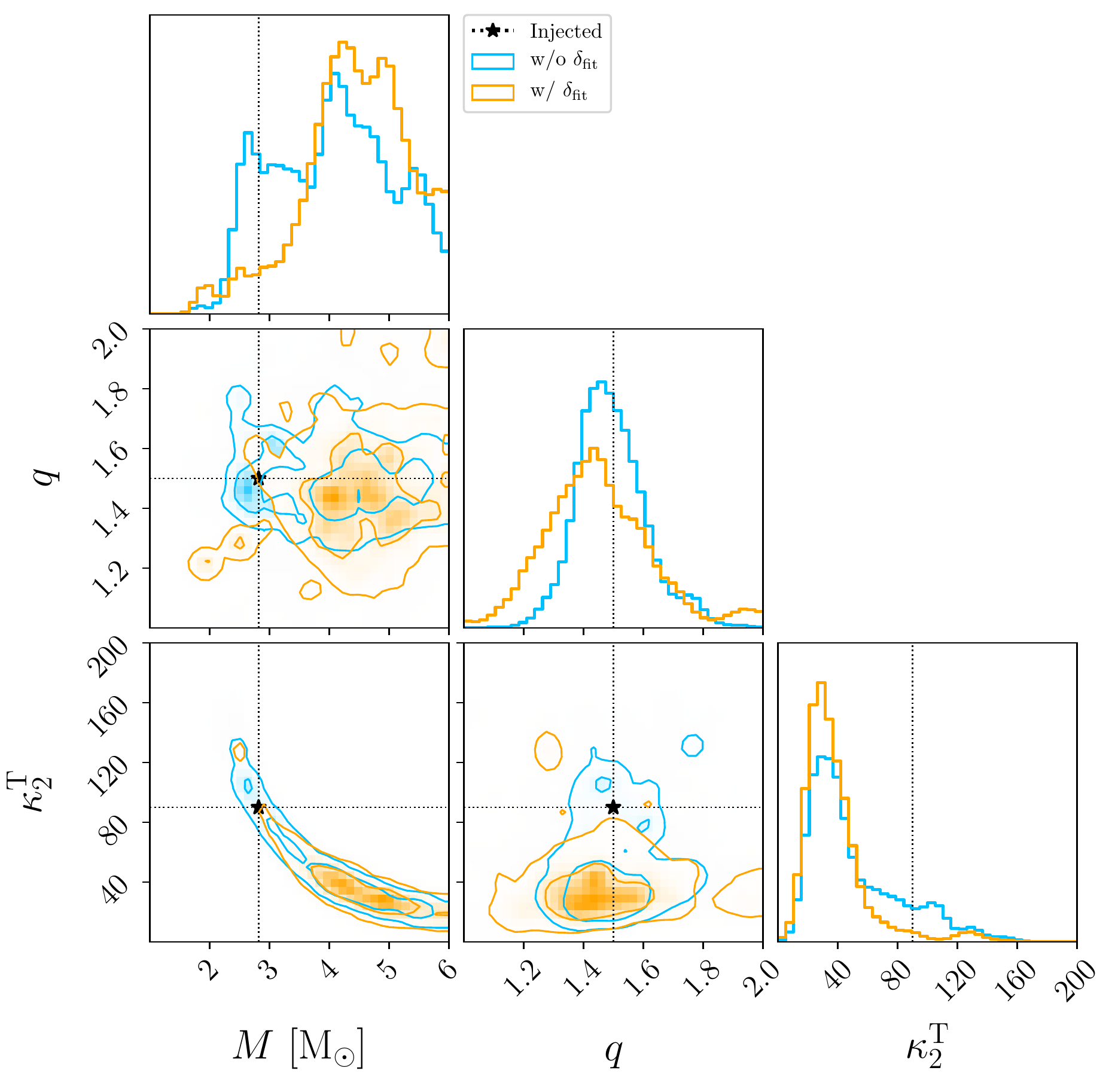}
	\includegraphics[width=0.49\textwidth]{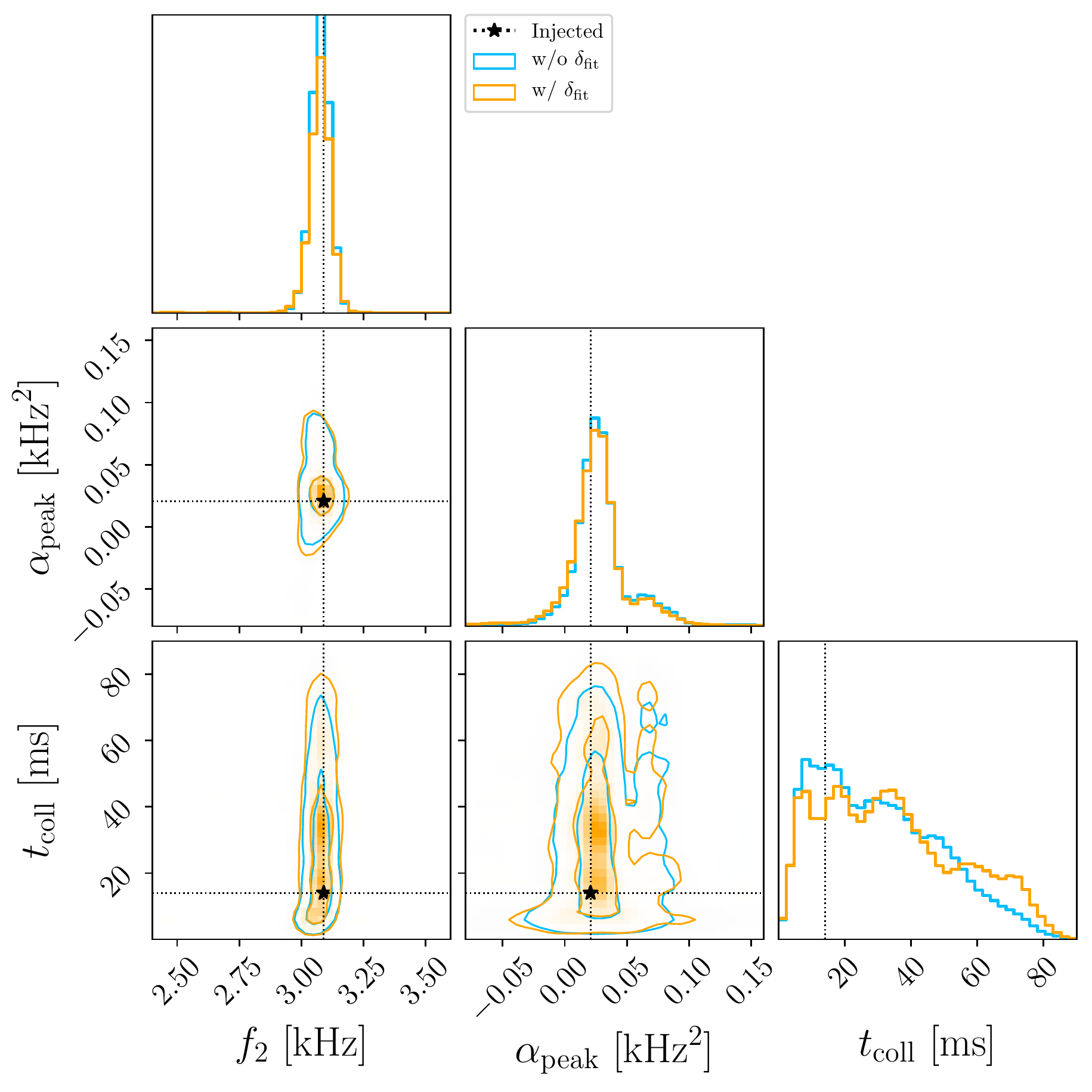}
	\caption{Posterior distributions for the 
		binary parameters $\{M,q,\kt\}$ (left)
		and the PM parameters 
		$\{f_2,\adrift, \tc\}$ (right)
		for the {\model} injection-recovery 
		study with SNR 8 (top) and SNR 10 (bottom).
				The reported posteriors 
		correspond to the average distributions over all employed noise realizations.
		The blue lines show the non-recalibrated
		posteriors; while, the orange lines 
		corresponds to the results that include
		the recalibrations $\recalib_{\rm fit}$.
		The contours show the 50\% and 90\%
		credibility regions.
		The injected values are reported with 
		black dotted lines.
		The total mass $M$
		and the tidal polarizability $\kt$
		are generally biased
		due to the correlations 
		introduced by the EOS-insensitive 
		relations.
	}
	\label{fig:nrpmw-post}
\end{figure*}

Figure~\ref{fig:nrpmw-post}
shows the posterior distributions
recovered for SNR 8 and 10.
The figures show the direct comparison
between the non-recalibrated and the 
recalibrated posteriors.
The results show
negligible differences in the recovered
parameters. 
The recalibrated inference
shows, as expected, a small broadening of the
posteriors. 
The recalibrated posteriors give a more conservative measurement
because they take into account the uncertainties of the quasiuniversal relations.

At the considered SNRs, we find that
the total mass is generally biased toward 
larger values, due 
to the correlations with
other parameters, in particular with
tidal polarizability $\kt$,
luminosity distance $D_L$
and time of collapse $\tc$.
Additional investigations with different
prior boundaries for $M$ show that these biases
persist also with larger upper bounds.
As a consequence, 
the recovered tidal polarizabilities $\kt$
typically underestimate the true value 
due to the analytical form of the 
EOS-insensitive relations,
as evident from the posterior
contours in the $\{M,\kt\}$ plane.
The posteriors slightly shrinks for increasing 
SNR, showing errors of ${\sim}1~\Msun$ and ${\sim}40$ respectively
for $M$ and $\kt$ at SNR 10
and 
making the biases more evident.
However, the injected values are included
within the $90\%$ confidence levels
up to SNR 10.
These biases can be fixed introducing
the pre-merger information in the PE
routine~\cite[e.g.][]{Breschi:2019srl,
	Wijngaarden:2022sah}.

On the other hand,
the mass ratio is well identified
with errors of ${\sim }0.3$ at PM SNR 8, 
since it strongly correlates with the ratio
of merger and PM amplitudes.
Also the posteriors on the luminosity distance $D_L$
are generally consistent with the injected values
with errors of $O(100~{\rm Mpc})$
at PM SNR 8.
The frequency drift $\adrift$ 
and time of collapse $\tc$
show informative measurements
that confidently include the injected values
within the posterior support.
However, 
these quantities are  poorly constrained 
with errors comparable to the injected values.
This is related to the nature of these terms:
the parameters 
$\{\adrift,\tc\}$
contribute to the late-time features of {\model} 
affecting the length and the frequency evolution
of the PM tail.
This portion of the PM signal is typically 
less luminous than the previous segment~\cite{Zappa:2017xba}
and noise contributions become more
relevant.

Focusing on the recalibrated PE,
the PM peak frequency $f_2$ 
is estimated to be 
$f_2={3.07}^{+0.04}_{-0.10}~{\rm kHz}$
for PM SNR 8
and 
$f_2={3.08}^{+0.02}_{-0.05}~{\rm kHz}$
for PM SNR 10,
where the errors correspond
to the 90\% confidence
level.
The corresponding recovered tidal polarizabilities 
are 
$\kt ={33}^{+105}_{-18}$ and 
$\kt ={32}^{+85}_{-14}$.
These results
improve on the estimates of 
the non-recalibrated analyses of
consistent injection-recovery performed with
{\oldmodel}~\cite{Breschi:2019srl,Breschi:2021wzr}.
The constrains on the frequency drift $\adrift$
improve from 
$\adrift ={0.032}^{+0.109}_{-0.155}~{\rm kHz}^2$
at PM SNR 8 to
$\adrift ={0.026}^{+0.047}_{-0.037}~{\rm kHz}^2$
at PM SNR 10.
On the other hand,
the estimates for the times of collapse 
$\tc$ do not report significant
improvement for the considered SNRs,
recovering
$\tc ={34}^{+37}_{-29}~{\rm ms}$
for PM SNR 8 and
$\tc ={32}^{+40}_{-27}~{\rm ms}$
for PM SNR 10.

\section{NR injections}
\label{sec:nr}

 \begin{figure*}[t]
	\centering 
	\includegraphics[width=0.99\textwidth]{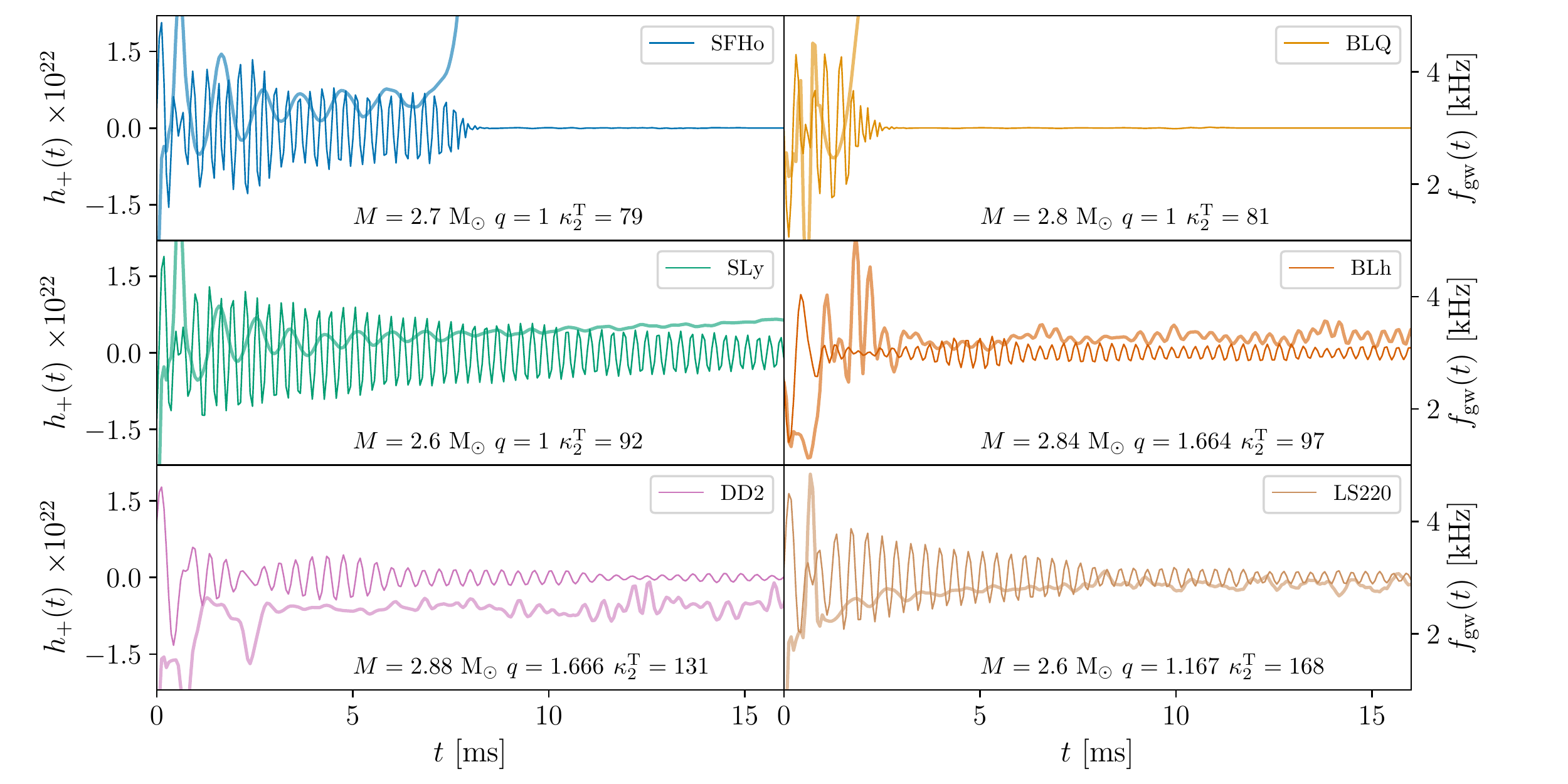}
	\caption{GW signals of the NR validation binaries
					locating the sources at $D_L=100~{\rm Mpc}$.
					The left axes show the plus polarization $h_{+}(t)$
					(thin solid lines)
					and the right axes show the instantaneous GW
					frequency $f_{\rm gw}(t)$
					(thick shadowed lines).
					The panels focus on the first $16~{\rm ms}$ after merger.
					The time of merger is fixed to $t_{\rm mrg}=0$ for each binary. 
					The signals are sampled with a rate of $16384~{\rm Hz}$.}
	\label{fig:wfs}
\end{figure*}

 \begin{table*}[t]
	\centering    
	\caption{Summary of the simulation study
		performed with {\model} on NR validation data.
		The first column reports the morphology 
		of the remnant, i.e.
		short-lived remnant, long-lived remnant or 
		tidally disrupted remnant.
		The following seven columns show the NR data properties,
		i.e. EOS, reference, total mass $M$, mass ratio $q$, tidal polarizability $\kt$, 
		PM peak frequency $f_2$ and the time of BH collapse $\tc$.
		The ninth column report the PM SNR $\rho_{\rm inj}$ of the injected NR data.
		The remaining columns indicate the median values and the 90\% credible regions
		of the recovered posterior distributions obtained 
		in the injection study averaged 
		over the different noise realizations.
		The medians and nominal errors for the $\log\mathcal{B}$ are also obtained  
		varying noise realizations
		and the nominal errors correspond 
		to the maximum and minimum recovered
		values.}
	\resizebox{\textwidth}{!}{
		\begin{tabular}{c|cccccccc|cccccc}        
			\hline
			\hline
			\multicolumn{9}{c|}{Injected properties}&\multicolumn{6}{c}{Recovered properties}\\
			\hline
             Morph. &EOS & Ref.&
             $M$ & $q$ & $\kt$& $f_2$& $\tc$
             &
             $\rho_{\rm inj}$& 
             $M$ & $q$ & $\kt$ & 
             $f_2$& 
             $\tc$&
             $\log\mathcal{B}$\\
                & & &
             $[\Msun]$   &   
             & &$[{\rm kHz}]$&$[{\rm ms}]$&
             & 
             $[\Msun]$ &  &   & 
             $[{\rm kHz}]$& 
             $[{\rm ms}]$& \\
             \hline
             \hline
			\multirow{6}{*}{\rotatebox[origin=c]{90}{Short-lived}} & 	
             \multirow{3}{*}{SFHo} & 
			\multirow{3}{*}{\cite{Bernuzzi:2015opx}}&
			\multirow{3}{*}{$2.7$} & 
			\multirow{3}{*}{$1.0$} & 
			\multirow{3}{*}{$79$}&
			\multirow{3}{*}{$3.42$}&
			\multirow{3}{*}{$8.0$}&
			7&${3.4}^{+2.2}_{-1.8}$&${1.33}^{+0.59}_{-0.31}$&${318}^{+376}_{-270}$&
			${2.0}^{+1.7}_{-0.9}$ &${21}^{+42}_{-18}$ 
			&${3.3}^{+1.6}_{-3.4}$  \\
			&&&&&&&&8&${3.7}^{+1.7}_{-1.7}$&${1.29}^{+0.60}_{-0.22}$&${37}^{+129}_{-29}$ &
			${3.4}^{+0.3}_{-2.0}$ &${29}^{+31}_{-24}$
			&${11.5}^{+1.6}_{-10.1}$ \\
			&&&&&&&&10&${4.3}^{+0.6}_{-1.9}$&${1.13}^{+0.23}_{-0.09}$&${21}^{+73}_{-18}$ &
			${3.41}^{+0.05}_{-0.04}$ &${30}^{+28}_{-17}$
			&${29.8}^{+1.6}_{-12.7}$ \\
			\cline{2-15}
            &  \multirow{3}{*}{BLQ} & 
             \multirow{3}{*}{\cite{Prakash:2021wpz}}&
             \multirow{3}{*}{$2.8$} & 
             \multirow{3}{*}{$1.0$} & 
             \multirow{3}{*}{$81$}&
             \multirow{3}{*}{$3.44$}&
             \multirow{3}{*}{$2.7$}&	
             7&${3.4}^{+2.0}_{-1.7}$&${1.27}^{+0.66}_{-0.24}$&${233}^{+474}_{-211}$&
             ${2.4}^{+0.7}_{-1.2}$&${23}^{+43}_{-20}$
             &${4.5}^{+1.5}_{-4.3}$ \\
             &&&&&&&&8&${2.9}^{+1.4}_{-1.2}$&${1.25}^{+0.72}_{-0.21}$&${83}^{+173}_{-73}$&
             ${2.6}^{+0.8}_{-1.1}$&${19}^{+26}_{-17}$
             &${10.4}^{+1.0}_{-5.2}$ \\
             &&&&&&&&10&${2.6}^{+2.4}_{-0.8}$&${1.17}^{+0.08}_{-0.15}$&${87}^{+174}_{-61}$&
             ${2.65}^{+0.61}_{-0.12}$&${2}^{+35}_{-2}$
             &${28.5}^{+0.9}_{-4.4}$ \\
          	\hline
          	\multirow{6}{*}{\rotatebox[origin=c]{90}{Long-lived}} & 	
         	\multirow{3}{*}{SLy} & 
             \multirow{3}{*}{\cite{Breschi:2019srl}}&
             \multirow{3}{*}{$2.6$} & 
             \multirow{3}{*}{$1.0$} & 
             \multirow{3}{*}{$92$}&
             \multirow{3}{*}{$3.13$}&
             \multirow{3}{*}{$21$}& 
			7&${4.1}^{+1.6}_{-2.2}$&${1.25}^{+0.59}_{-0.20}$&${217}^{+467}_{-199}$ &
			${2.2}^{+1.3}_{-1.1}$ &${34}^{+29}_{-29}$ 
			&${2.1}^{+1.5}_{-2.3}$ \\
			&&&&&&&&8&${4.4}^{+1.3}_{-2.3}$&${1.45}^{+0.48}_{-0.36}$&${40}^{+217}_{-33}$ &
			${3.2}^{+0.3}_{-1.9}$ &${31}^{+33}_{-22}$ 
			&${8.0}^{+1.6}_{-8.2}$ \\
			&&&&&&&&10&${3.4}^{+1.8}_{-0.8}$&${1.20}^{+0.26}_{-0.14}$&${46}^{+58}_{-34}$ &
			${3.26}^{+0.06}_{-0.07}$ &${30}^{+33}_{-21}$ 
			&${25.5}^{+1.6}_{-9.6}$ \\
             \cline{2-15}
             &\multirow{3}{*}{LS220} & 
             \multirow{3}{*}{\cite{Perego:2019adq}}&
             \multirow{3}{*}{$2.6$} & 
             \multirow{3}{*}{$1.167$} & 
             \multirow{3}{*}{$168$}&
             \multirow{3}{*}{$2.68$}&
             \multirow{3}{*}{$35$}&
			7&${3.7}^{+2.0}_{-1.5}$&${1.23}^{+0.37}_{-0.22}$&${67}^{+227}_{-55}$&
			${2.7}^{+0.2}_{-0.2}$&${25}^{+34}_{-18}$
			&${6.4}^{+1.3}_{-3.7}$ \\
			&&&&&&&&8&${3.9}^{+1.8}_{-1.3}$&${1.30}^{+0.30}_{-0.26}$&${55}^{+95}_{-41}$&
			${2.67}^{+0.15}_{-0.13}$&${17}^{+27}_{-13}$
			&${13.9}^{+1.2}_{-4.8}$ \\
			&&&&&&&&10&${3.4}^{+2.3}_{-1.5}$&${1.23}^{+0.32}_{-0.18}$&${70}^{+126}_{-62}$&
			${2.68}^{+0.12}_{-0.11}$&${18}^{+25}_{-10}$
			&${33.9}^{+1.4}_{-5.6}$ \\
			\hline
			\multirow{6}{*}{\rotatebox[origin=c]{90}{Tidally disrupted}} & 	
			\multirow{3}{*}{BLh} & 
			\multirow{3}{*}{\cite{Bernuzzi:2020txg}}&
			\multirow{3}{*}{$2.84$} & 
			\multirow{3}{*}{$1.664$} & 
			\multirow{3}{*}{$97$}&
			\multirow{3}{*}{$3.27$}&
			\multirow{3}{*}{$18$}&
			7&${3.1}^{+2.1}_{-1.2}$&${1.18}^{+0.77}_{-0.17}$&${503}^{+257}_{-436}$&
			${1.6}^{+1.6}_{-0.7}$&${20}^{+27}_{-18}$
			&${8.9}^{+1.5}_{-5.3}$ \\
			&&&&&&&&8&${3.2}^{+2.2}_{-1.0}$&${1.17}^{+0.76}_{-0.15}$&${87}^{+195}_{-78}$&
			${1.5}^{+2.5}_{-0.6}$&${28}^{+24}_{-25}$
			&${16.6}^{+1.5}_{-5.7}$ \\
			&&&&&&&&10&${3.3}^{+2.3}_{-0.7}$&${1.12}^{+0.81}_{-0.11}$&${62}^{+216}_{-57}$&
			${1.5}^{+1.9}_{-0.8}$&${19}^{+32}_{-18}$
			&${38.2}^{+1.6}_{-7.8}$ \\
			\cline{2-15}
		   &\multirow{3}{*}{DD2} & 
			\multirow{3}{*}{$\substack{\text{This} \\ \text{work}}$}&	
		   \multirow{3}{*}{$2.88$} & 
		   \multirow{3}{*}{$1.666$} & 
		   \multirow{3}{*}{$131$}&
		   \multirow{3}{*}{$2.58$}&
		   \multirow{3}{*}{$34$}&	 
			7&${3.8}^{+1.8}_{-1.5}$&${1.41}^{+0.52}_{-0.38}$&${261}^{+432}_{-240}$&
			${1.9}^{+0.9}_{-0.8}$&${20}^{+42}_{-18}$
			&${5.8}^{+1.5}_{-5.2}$ \\					
			&&&&&&&&8&${3.5}^{+1.8}_{-1.4}$&${1.50}^{+0.42}_{-0.47}$&${92}^{+172}_{-73}$&
			${2.2}^{+0.5}_{-1.1}$&${22}^{+49}_{-20}$
			&${14.5}^{+1.6}_{-9.0}$ \\
			&&&&&&&&10&${3.3}^{+1.7}_{-1.2}$&${1.63}^{+0.23}_{-0.59}$&${86}^{+130}_{-63}$&
			${2.13}^{+0.54}_{-0.95}$&${9}^{+31}_{-8}$
			&${31.8}^{+1.6}_{-12.2}$ \\
			\hline
			\hline
	\end{tabular}}
	\label{tab:inject}
\end{table*}

In this section, we perform full
  PE injection-recovery experiments using NR signals and {\model}; the
  NR validation set and the main results are summarized in 
  Table~\ref{tab:inject}.
The validation set is composed by six non-spinning binaries
computed with {\scshape THC}~\cite{Radice:2012cu}
which simulate
microphysics, neutrino transport (with various schemes) 
and turbulent viscosity.
The set includes
two long-lived remnants 
(SLy with total mass $1.30{+}1.30~\Msun$~\cite{Breschi:2019srl} 
and LS220 $1.40{+}1.20~\Msun$~\cite{Perego:2019adq}),
two short-lived remnants 
(SFHo $1.35{+}1.35~\Msun$~\cite{Bernuzzi:2015opx} 
and BLQ $1.40{+}1.40~\Msun$~\cite{Prakash:2021wpz}),
and two large-mass-ratio binaries with
tidal disruptive morphology (DD2 $1.80{+}1.08~\Msun$~\cite{Breschi:2022}
and BLh $1.772{+}1.065~\Msun$~\cite{Bernuzzi:2020txg}).

Figure~\ref{fig:wfs} shows the corresponding GW signals from NR simulations.
The short-lived SFHo $1.35{+}1.35~\Msun$ remnant
shows prominent modulations around the carrier 
frequency $f_2\simeq 3.4~{\rm kHz}$
and increasing frequency drift 
corresponding to $\adrift\simeq 0.075~{\rm kHz}^2$.
The remnant collapses at
$\tc\simeq 8~{\rm ms}$.
The second short-lived case,
BLQ $1.40{+}1.40~\Msun$,
is computed with an EOS that includes
a deconfined quark phase.
The remnant collapses into a BH shortly after the first bounce 
of the core collision, at $\tc\simeq 3~{\rm ms}$,
and the corresponding $f_2\simeq  3.4~{\rm kHz}$ 
peak deviates from the 
quasiuniversal relation above the $90\%$ credibility region (see {\paperI}).
Moreover, for this binary, it is not possible to estimate 
the frequency drift parameter $\adrift$
since the remnant collapses before the third nodal point (see
{\paperI}, for the definition of $\adrift$ and nodal points).
The long-lived SLy $1.30{+}1.30~\Msun$ remnant shows
prominent modulations around the carrier 
frequency $f_2\simeq 3.1~{\rm kHz}$
with $\adrift\simeq 0.032~{\rm kHz}^2$
and collapses into a BH $21~{\rm ms}$ after merger.
The second long-lived remnant,
LS220 $1.40{+}1.20~\Msun$,
survives $35~{\rm ms}$ after merger,
and its spectrum peaks at $f_2\simeq 2.7~{\rm kHz}$.
For times $\lesssim 20~{\rm ms}$ after merger, 
the corresponding GW PM signal
shows an increasing frequency evolution with $\adrift\simeq 0.027~{\rm kHz}^2$;
however, the instantaneous frequency decays during the final stages
with $\adrift\simeq -0.037~{\rm kHz}^2$.
Averaging over the entire PM duration,
we estimate an $\adrift$ of $-0.01~{\rm kHz}^2$.
The tidal disruptive BLh $1.772{+}1.065~\Msun$ remnant 
collapses into a BH $18~{\rm ms}$ after merger, and it  
corresponds to the intrinsically fainter PM GW signal of our set.
The frequency evolution is milder compared to the previous cases,
with $\adrift\simeq 0.015~{\rm kHz}^2$ and its spectrum
shows $f_2\simeq3.3~{\rm kHz}$ with multiple subdominant peaks.
The second tidal disruptive case,
DD2 $1.80{+}1.08~\Msun$,
generates a remnant that collapses $34~{\rm ms}$
after merger and it approximately shows a constant
frequency evolution, $\adrift\simeq  5 {\times }10^{-3}~{\rm kHz}^2$.
The corresponding spectrum shows 
PM peak centered 
around ${\sim}2.6~{\rm kHz}$.

In Sec.~\ref{sec:detect}
we investigate the signal detectability
as a function of the injected SNR of the NR data.
We discuss the posterior distributions for the 
spectra in Sec.~\ref{sec:spectra}
and for the characteristic PM frequencies
in Sec.~\ref{sec:fpeak}.
Then, we present the obtained constraints on binary parameters (Sec.~\ref{sec:binparams})
and the PM parameters (Sec.~\ref{sec:pmparams}).

\subsection{Detectability}
\label{sec:detect}

 \begin{figure*}[t]
	\centering 
	\includegraphics[width=0.99\textwidth]{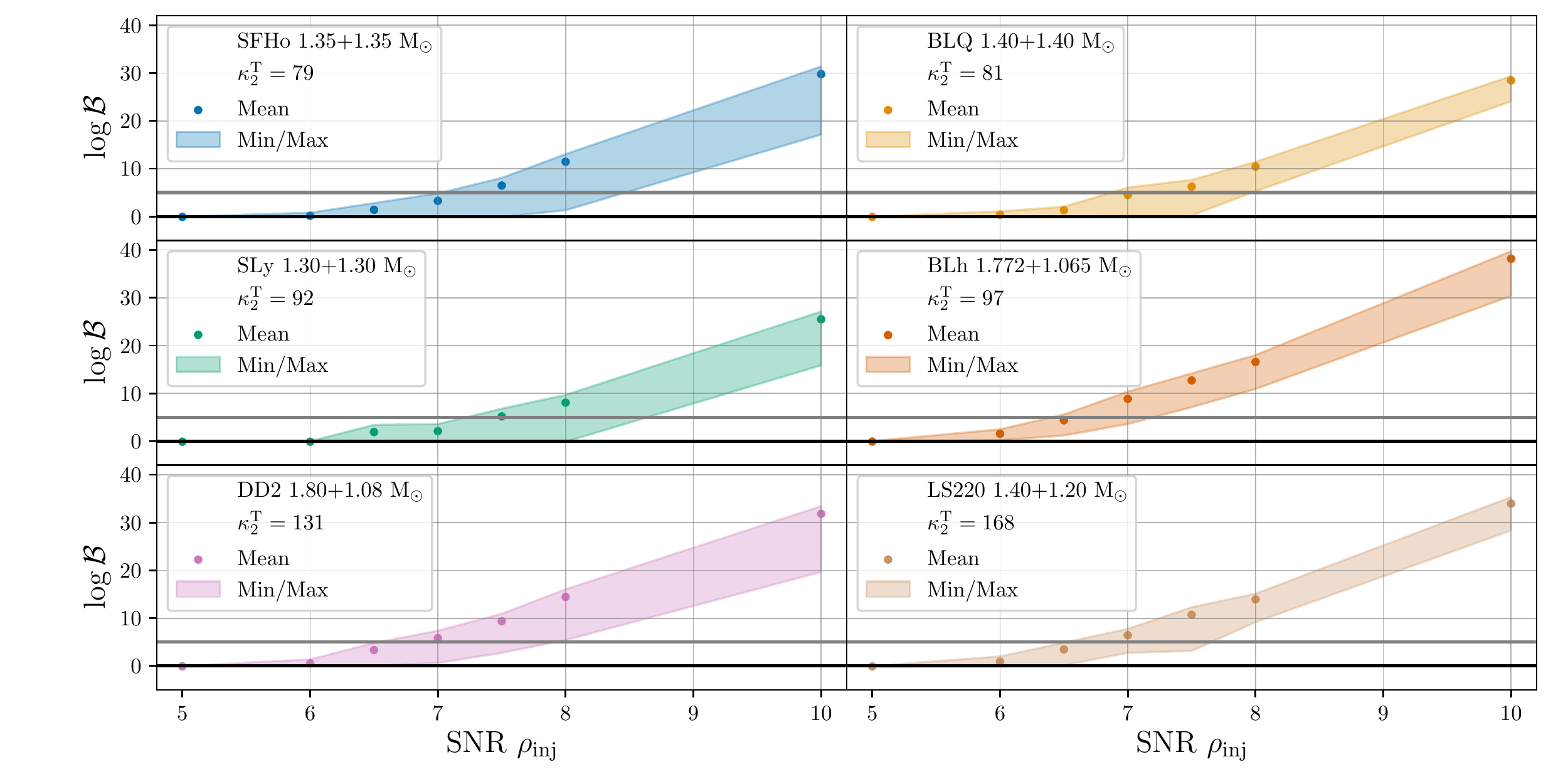}
	\caption{Logarithmic BFs $\log\B$ as functions of the PM SNR 
					$\rho_{\rm inj}$ of the injected NR
					template from Table~\ref{tab:inject}.
					The dots refer to the mean values
					averaged over the different noise realizations
					and the shadowed areas correspond to the 
					minimum and maximum values recovered in 
					the survey.
					Two horizontal lines identify $\log\B=0$ (black) and 
					$\log\B=5$ (gray).}
	\label{fig:bfs}
\end{figure*}

Figure~\ref{fig:bfs} shows 
the recovered mean, maximum and minimum BFs 
for the binaries listed in Table~\ref{tab:inject}. 
In general, 
the BFs show the expected increasing trend
deviating from $\log\B \simeq 0$ 
and recovering informative posterior distributions
at PM SNR $\rho_{\rm inj}\simeq 6$.
Averaging over all the analyzed cases,
the nominal detectability threshold 
is reached for PM SNR $\rho_{\rm inj} = 7.2^{+0.8}_{-0.5}$.
Noise fluctuations affect the 
estimates, leading to 
larger threshold SNRs
similarly to the analysis of Sec.~\ref{sec:nrpmw-detect},
but with more pronounced 
and less homogeneous fluctuations.
The recovered BFs
for the binaries with $\kt \lesssim 90$ 
have a slower trends
compared with the other cases and 
the case-study of Sec.~\ref{sec:template},
showing $\log\B \lesssim 30$ for SNR 10.
The corresponding PM 
transients show the most significant modulation
effects among the considered cases
and a short duration, except for SLy $1.30{+}1.30~\Msun$.
Moreover, the characteristic PM peaks 
for these cases are located at higher frequency values compared to the
other binaries, i.e.
$f_2 \gtrsim 3~{\rm kHz}$,
where the noise contributions increase.
The detectability threshold for these cases could be improved
by a refined characterization of the frequency modulations
(e.g. introducing a free modulation phase as a free parameter), or
refining the late-time portion of the template
(e.g. including the wavelet for the BH collapse).

\subsection{Spectra}
\label{sec:spectra}

 \begin{figure*}[t]
	\centering 
	\includegraphics[width=0.99\textwidth]{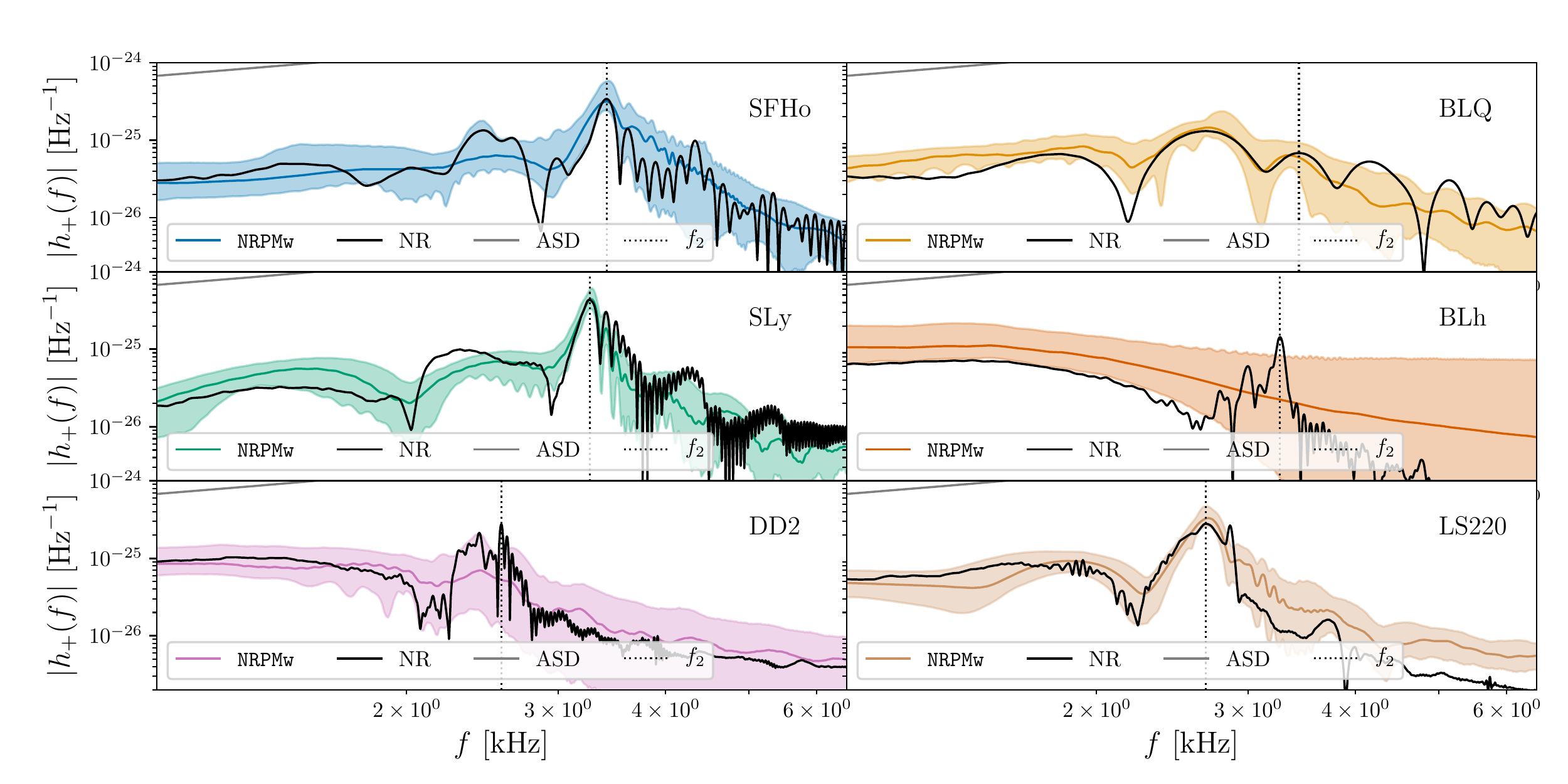}
	\caption{Posterior distributions of the GW spectra
					for the PMs plus polarization $h_+(f)$
					recovered for PM SNR $\rho_{\rm inj}=10$
					for an illustrative noise realization.
					The colored lines report the medians
					and the shadowed regions correspond
					to the 90\% credibility region.
					The injected spectra are reported with 
					black solid lines
					and the corresponding $f_2$ peaks
					are denoted with vertical black dotted lines.}
	\label{fig:sps}
\end{figure*}

Figure~\ref{fig:sps} shows 
the posterior distribution
of the recovered spectra for 
an illustrative 
noise realization at $\rho_{\rm inj}=10$.
We opt to report this case because 
it corresponds to the loudest employed SNR
and therefore the systematic errors between
{\model} model and NR data are more evident.
In general, the majority of the injected signals 
are included within the 90\% confidence levels 
of the recovered spectra, consistently with the 
discussion on the recovered SNRs of Sec.~\ref{sec:detect}.

The large mass ratio binaries BLh and DD2 
underestimate the characteristic PM peak.
In particular, the predicted spectra for the
BLh case 
is primarily informed by the merger portion of data.
and it strongly deviates from the 
injected value
due to the faint PM burst, 
that does not permit a clear identification of the
dominant PM peak.
The 90\% confidence level of the 
recovered time-domain waveform shows 
a non-vanishing tail for late times; however, the
median value is consistent with zero, 
showing that the signal is not resolved.
These errors can be related with a non-optimal
modeling for large mass ratios, i.e. $q>1.5$.
In particular,
tidally-disruptive mergers 
show additional phase discontinuities
and multiple bumps in the time-domain GW
amplitude, related with the remnant dynamics.

\subsection{PM frequencies}
\label{sec:fpeak}

\begin{figure*}[t]
	\centering 
	\includegraphics[width=0.99\textwidth]{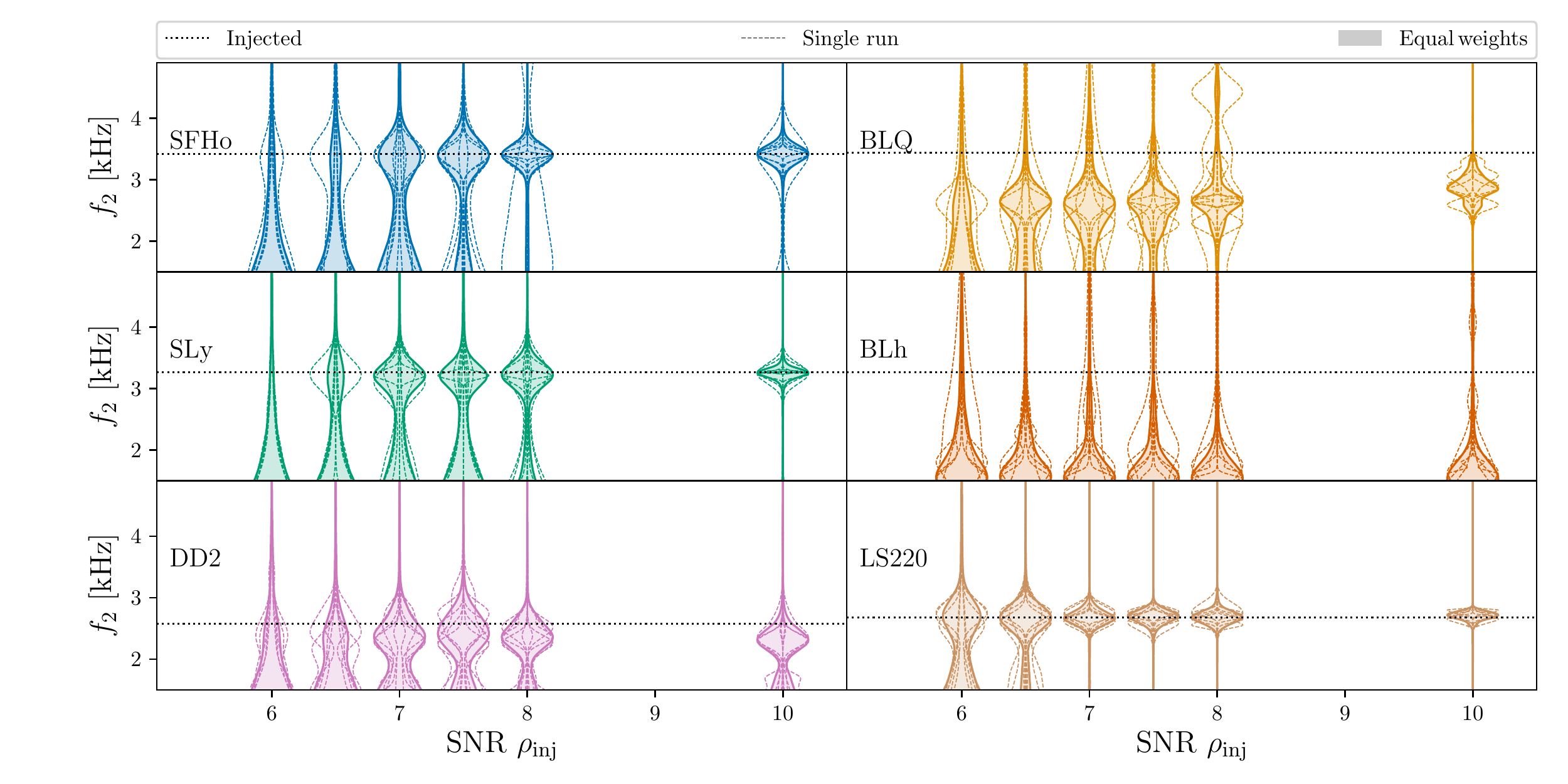}\\
	\includegraphics[width=0.99\textwidth]{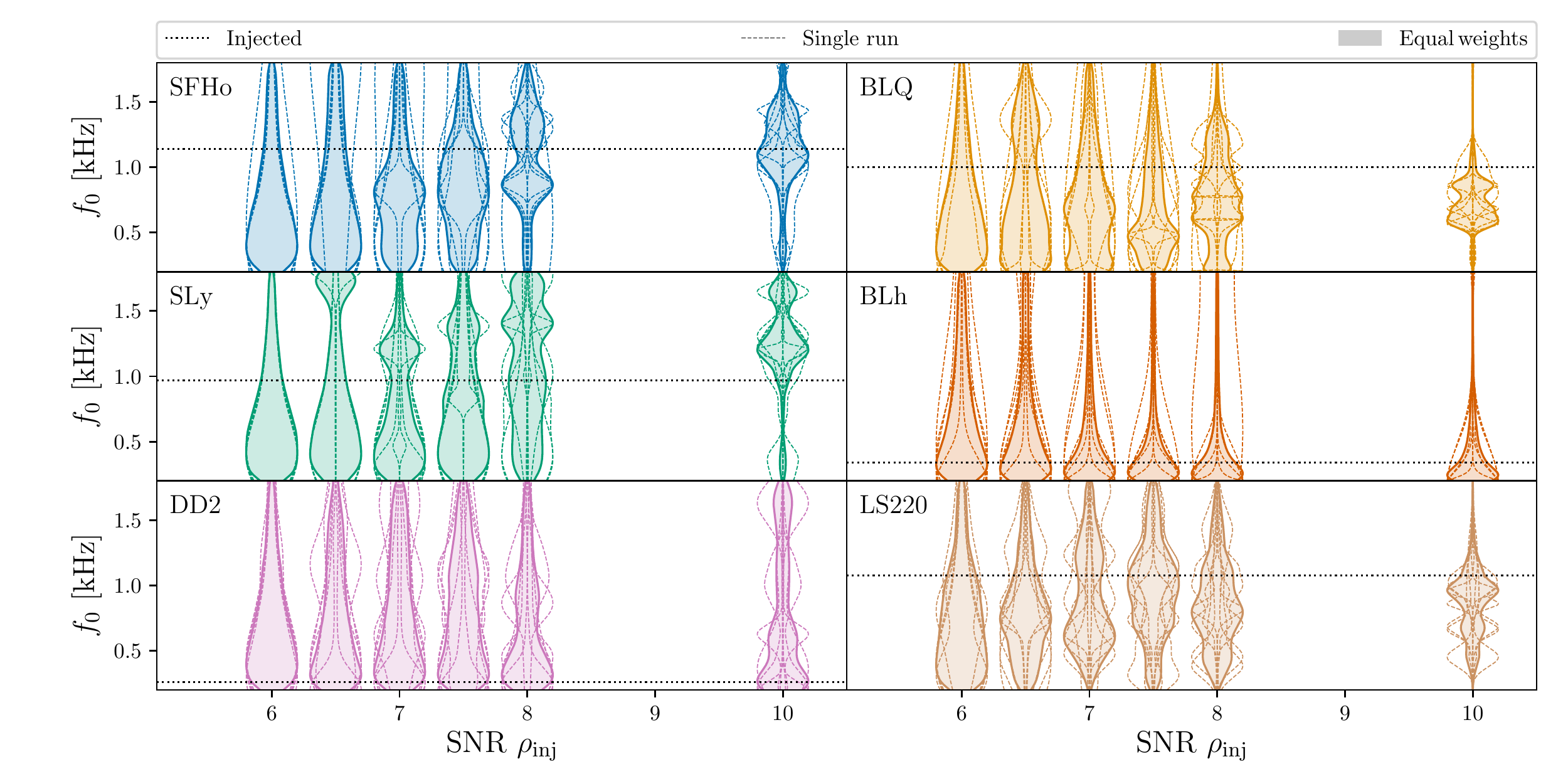}
	\caption{Posterior distributions for the PM  frequencies $f_2$ (top panel)
		and $f_0$ (bottom panel) 
		at the different PM SNR $\rho_{\rm inj}$
		recovered in the PE survey
		reported in Tab.~\ref{tab:inject}.
		For each case,
		thin dashed lines report the recovered posteriors for 
		each run and
		the filled areas show the equally-weighted combined 
		posterior distributions.
		Black horizontal dotted lines indicate the values
		of the injected NR templates.
	}
	\label{fig:f2s}
\end{figure*}

In {\model}, the actual $f_2$ value is 
determined by several quantities.
A first estimate relies on
the binary parameters through 
the quasiuniversal relation.
Then, 
the frequency drift parameter $\adrift$
and the value of the respective recalibrations $\recalibpm$ can shift the actual peak
from the prediction of the quasiuniversal relation.
Thus, a robust 
determination of the 
PM frequency can 
be estimated from the 
peak of the reconstructed spectra,
consistently with 
the extraction method discussed
in Sec.~{4 B} of {\paperI}.
For each sample, we generate the 
corresponding GW spectrum 
$h_{\model}(f)$,
considering only the time support $t>t_0$
in order to isolate the PM contribution of 
interest.
Then, $f_2$ is identified as the 
(typically dominant) spectral peak
of the carrier frequency component.
When the template corresponds to prompt BH
collapse,
a prior sample is extracted.

Figure~\ref{fig:f2s} (top panel) 
shows the recovered posterior distributions on $f_2$
as a function of the injected 
SNR $\rho_{\rm inj}$.
As mentioned in the previous section, for the majority of the cases
the recovered posteriors report informative inferences for 
$\rho_{\rm inj}\gtrsim 6$ with errors of $O(1~{\rm kHz})$.
The errors on the estimated $f_2$
decrease to $O(100~{\rm Hz})$
at PM SNR 10.
These results are comparable with 
similar estimates performed with unmodeled studies~\cite{Chatziioannou:2017ixj}
and strongly improve the results coming 
from damped-sinusoidal templates~\cite{Tsang:2019esi,Easter:2020ifj}.
Template-based analyses 
are expected to deliver better detectability
threshold than model-independent
estimates, if the employed template is a 
good representation of the signal.
Thus, in this perspective, this claim might appear counterintuitive.
However, template-based studies of 
PM transients
are generally showing poorer results than the more flexible models due to 
the considerable mismatch between the 
employed template and the signal.
In {\model}, these biases are corrected
with the recalibration $\recalib_{\rm fit}$,
accounting for deviations from the predictions of the calibrated relations.
At SNR $\rho_{\rm inj}=8$, all posterior distributions show 
informative measurements, with uncertainties ranging from $1~{\rm kHz}$ 
to $500~{\rm Hz}$, and the injected NR values are included in the $90\%$
credible regions.
The uncertainties on the $f_2$ posteriors reach $O(100~{\rm Hz})$
at SNR $\rho_{\rm inj}=10$ and systematic uncertainties start to play a more significant role.
In particular,
the analyses of the short-lived remnant
BLQ show bimodalities in the $f_2$ 
posteriors for some noise realizations.
These biases are attributable 
to subdominant couplings that have a larger contribution on the overall spectrum
for short-lived remnants,
whose power can exceed that of the $f_2$ peak
(see Fig.~\ref{fig:sps}).
Similar biases are recovered also in the analysis of the DD2
binary.
On the other hand, as discussed in Sec.~\ref{sec:spectra},
the inference of the BLh binary is not capable to
resolve the $f_2$ peak.

Another interesting quantity is the 
coupling frequency $f_0$.
This component induces a modulation
in of the carrier frequency $f_2$, generating
the subdominant spectral peaks, 
especially for very compact remnants .
The $f_0$ frequency is generally related 
with mass-density oscillations of the NS
remnant; thus, it can provide important 
insights on the NS structure~\citep[e.g.][]{Stergioulas:2003ep,
Bernuzzi:2008fu,Bernuzzi:2009sak,
Radice:2010rw,Stergioulas:2011gd}.
We estimate the $f_0$ posterior 
from the instrinsic binary properties
employing the quasiuniversal 
relations and modifying the prediction
with the associated recalibration parameter $\delta_{f_0}$.
Differently from $f_2$, the actual values of
$f_0$ is fully determined by these quantities
since {\model} assumes this frequency component
to be constant. 
Figure~\ref{fig:f2s} (bottom panel) 
shows the recovered posterior distributions on $f_0$
as a function of the injected 
SNR $\rho_{\rm inj}$.
The recovered $f_0$ posteriors are generally
broader than the $f_2$ ones, due to the weaker
magnitude of these spectral peaks.
The results show informative measurements
for PM SNR $\gtrsim 7.5$ with errors of 
$O(1~{\rm kHz})$. The uncertainties decrease
to $O(500~{\rm Hz})$ at PM SNR 10.
Moreover, noise fluctuations affect $f_0$ posterior
introducing biases and multimodalities
due to the weakness of the associated spectral peaks.

\subsection{Binary parameters}
\label{sec:binparams}

 \begin{figure*}[t]
	\centering 
		\includegraphics[width=0.99\textwidth]{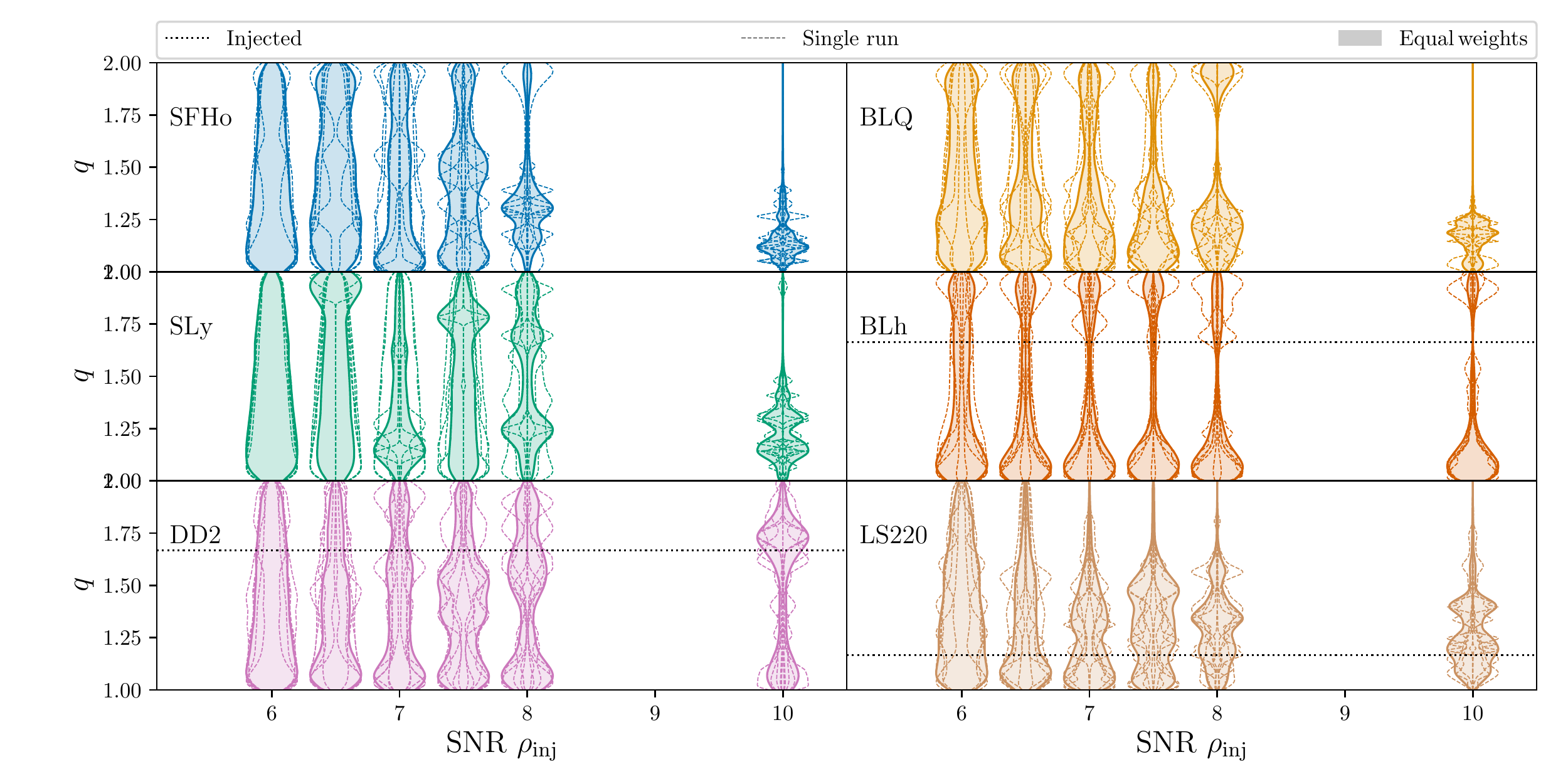}\\
	\includegraphics[width=0.99\textwidth]{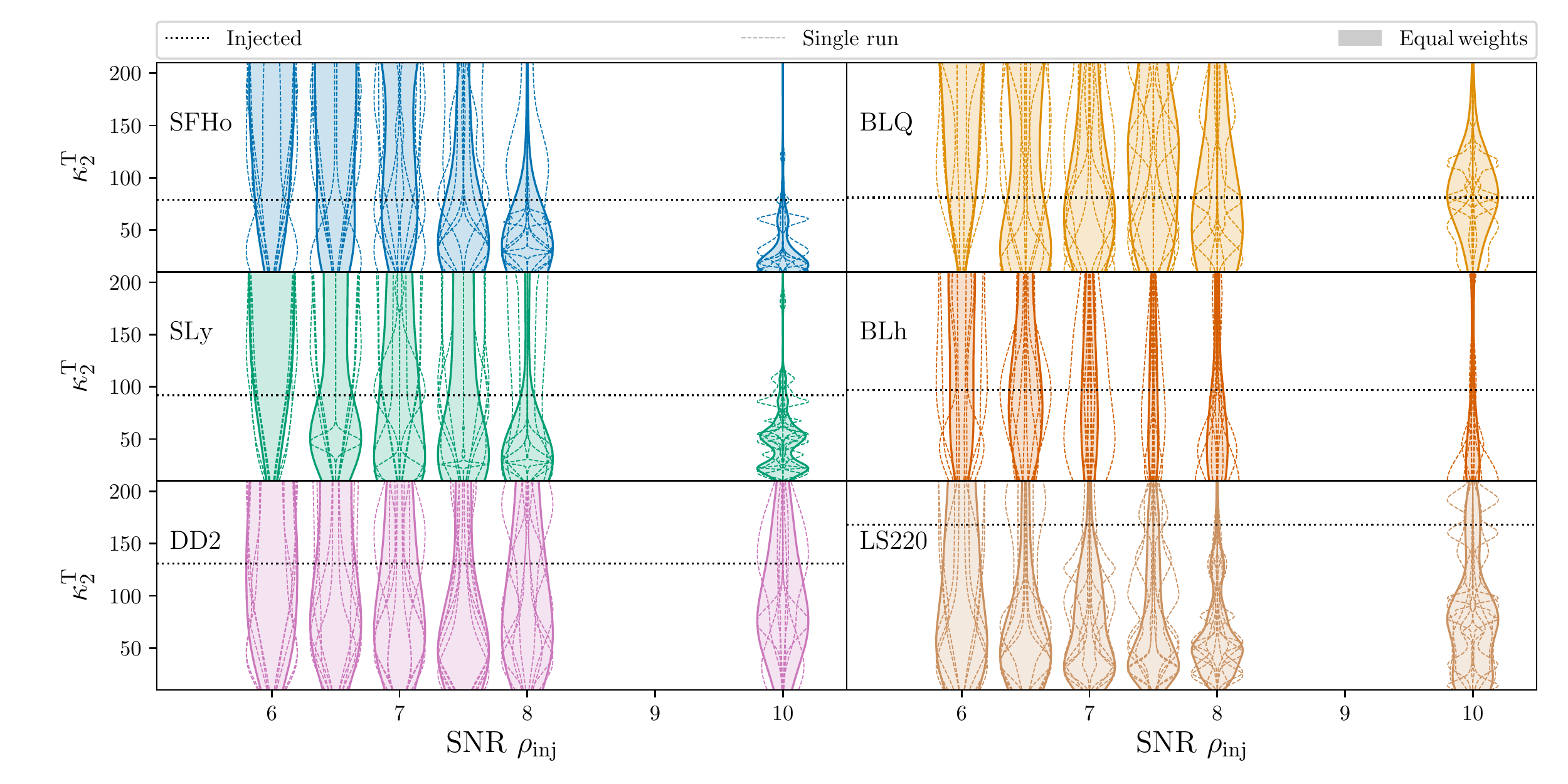}
	\caption{Posterior distributions for the 
		mass ratio $q$ (top panel) and the
		tidal polarizability $\kt$ (bottom panel)
					 at the different PM SNR $\rho_{\rm inj}$
					 recovered in the PE survey
					 reported in Tab.~\ref{tab:inject}.
					 For each case,
					 thin dashed lines report the recovered posteriors for 
					 each run and
					 the filled areas show the equally-weighted combined 
					 posterior distributions.
					 Black horizontal dotted lines indicate the values
					 of the injected NR templates.
		 		}
	\label{fig:k2s}
\end{figure*}

The intrinsic binary parameter are generally poorly constrained
compared to the pre-merger analysis due to the 
considerably smaller SNR of the PM signal.
Overall, the recovered posteriors show
a behavior similar to the case discussed 
in Sec.~\ref{sec:template}.
The median binary masses $M$ are 
typically shifted toward larger values,
with errors of $O(1~\Msun)$.
Nevertheless, the injected values are always included in the $90\%$
credible regions, indicating that our inference is unbiased at the SNRs under consideration.
The mass ratio,
shown in Fig.~\ref{fig:k2s} (top panel), 
is typically well 
identified up to PM SNR ${\sim}8$ 
with errors of ${\sim}0.6$
at sensitivity threshold.
For increasing SNRs, 
systematic errors become more dominant
especially for equal-mass binaries,
where the overall PM power is larger.
The tidal polarizability $\kt$
shows systematic errors for 
PM SNR$~\gtrsim 8$, underestimating 
the injected values, consistently
with the mass biases and analogously
to the studies in Sec.~\ref{sec:template}.
However, 
as shown in Fig.~\ref{fig:k2s} (bottom panel),
the average over 
the different noise realizations mitigates
the systematic biases,
including the injected values 
within the 90\% credibility
region for all considered binaries.
The spins are generally dominated 
by the prior up to PM SNR ${\sim }10$.

Comparing these results with the ones in Ref.~\cite{Breschi:2019srl},
we observe an overall widening of the posterior distributions,
which is related to the inclusion of the recalibrations $\recalibpm$.
However, at the same time the systematic biases are considerably reduced,
showing that such technique is able to providing accurate and conservative results.
The obtained results are consistent with the estimates of Ref.~\cite{Breschi:2021xrx}.

\subsection{PM parameters}
\label{sec:pmparams}

 \begin{figure*}[t]
	\centering 
	\includegraphics[width=0.99\textwidth]{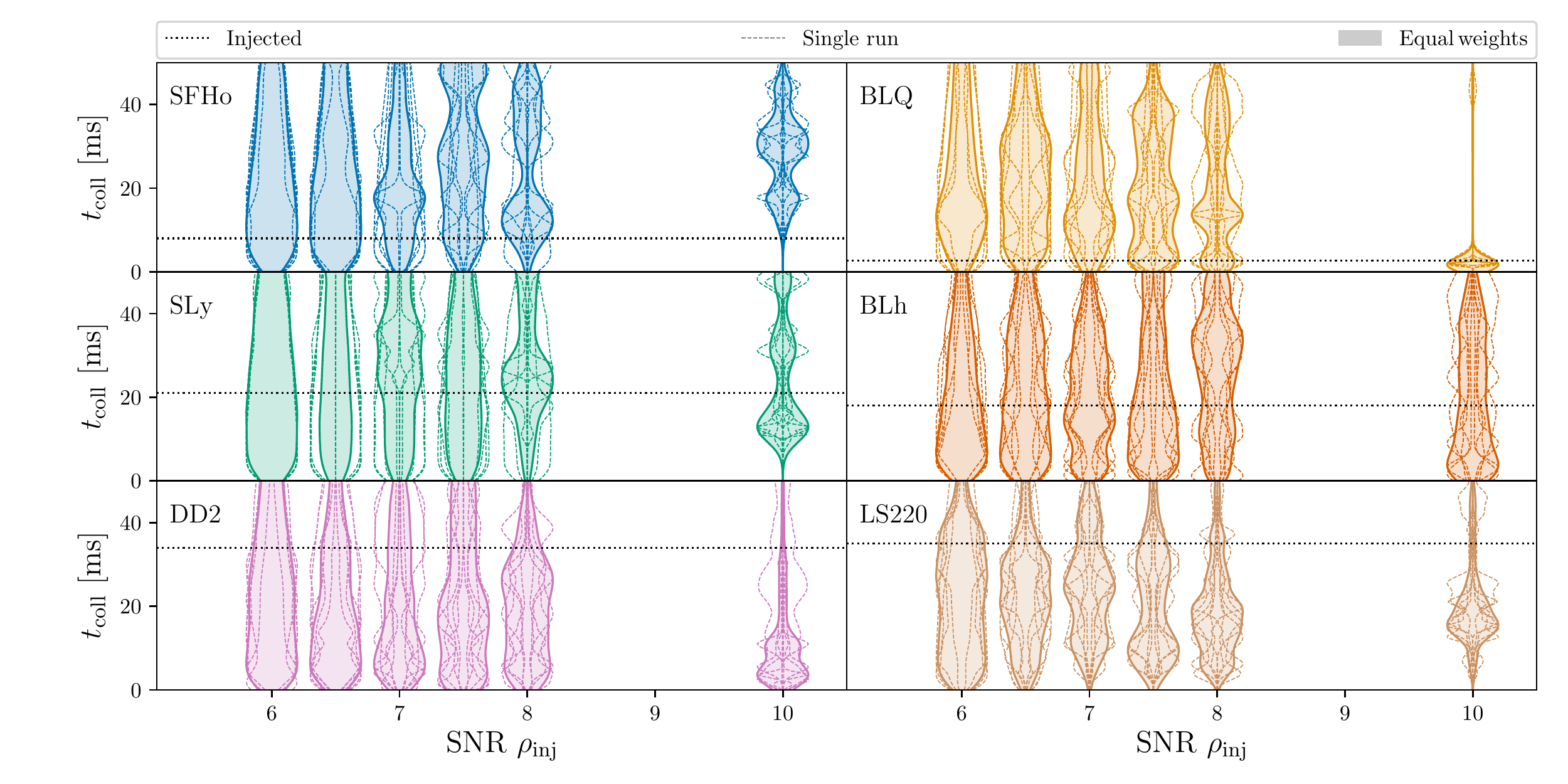}
	\includegraphics[width=0.99\textwidth]{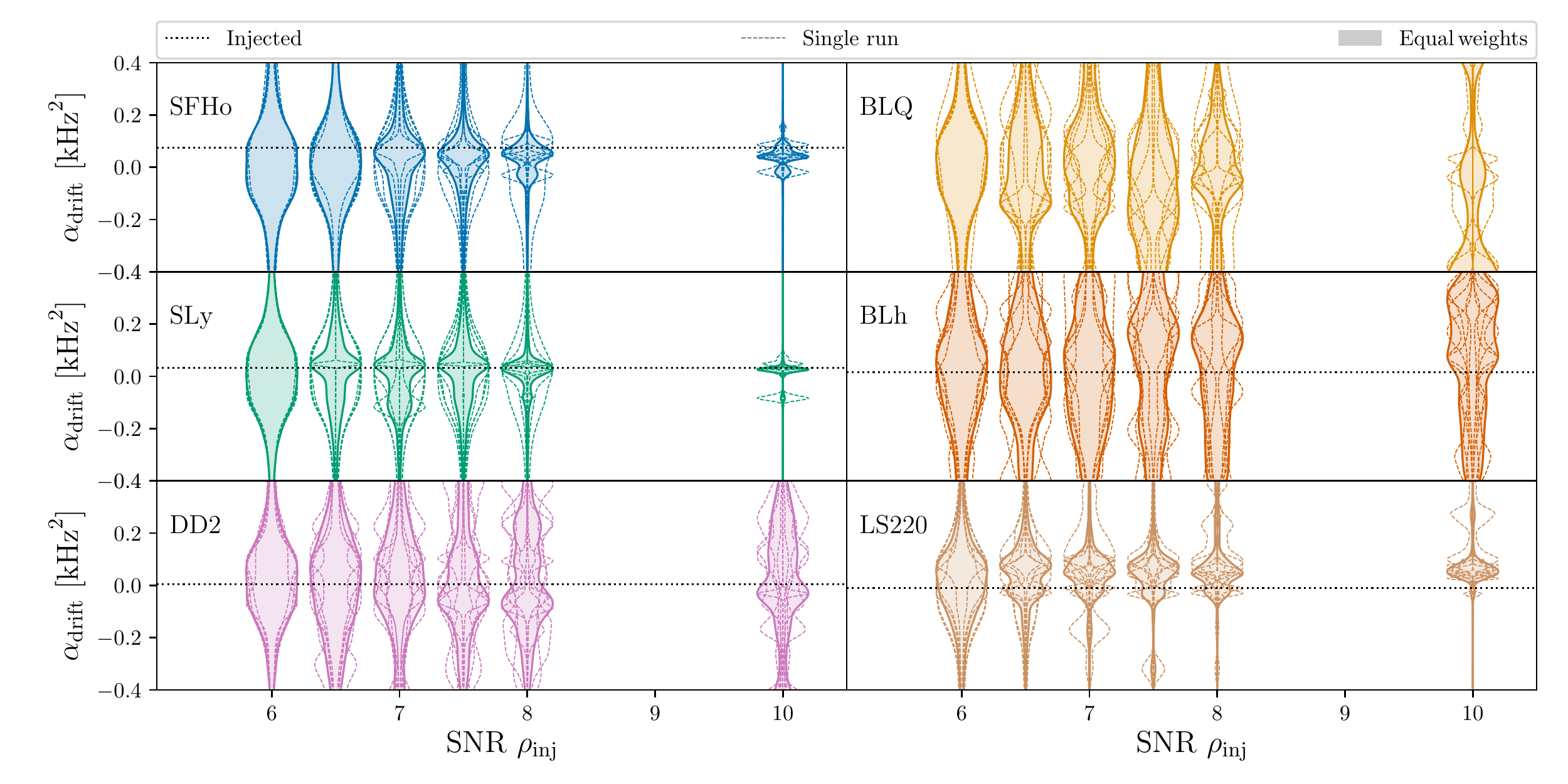}
	\caption{Posterior distributions for the time of 
		collapse $\tc$ (top panel)
		and the frequency drift $\adrift$ (bottom
		panel)
		at the different PM SNR $\rho_{\rm inj}$
		recovered in the PE survey
		reported in Tab.~\ref{tab:inject}.
		For each case,
		thin dashed lines report the recovered posteriors for 
		each run and
		the filled areas show the equally-weighted combined 
		posterior distributions.
		Black horizontal dotted lines indicate the values
		of the injected NR templates.
		In the BLQ case, extracting a prediction was not possible (see main text).}
	\label{fig:pms}
\end{figure*}

Figure~\ref{fig:pms} (top panel)
shows the posterior distributions
for the time of BH collapse $\tc$
as functions of the injected SNR.
This term is strongly affected by 
noise fluctuations, since the
late-time signal tail is no more observable
when its amplitude goes below the noise 
threshold.
However, the injected values are generally
included within the 90\% confidence levels.
The associated uncertainties go from 
$O(30~{\rm ms})$ at threshold SNR
to $O(20~{\rm ms})$ at PM SNR 10.
A particular case is the BLQ binary
that corresponds to the most massive
binary with the shortest PM transient 
among the considered cases.
The corresponding
posterior distribution for SNR 10
is tightly constrained around the injected 
value, implying that $\tc$
can be better estimated for very-short-lived
remnants.
The inference of $\tc$ can be improved 
introducing the model for the remnant collapse
in {\model} (see {\paperI}).
However, the observation of the BH collapse
is strongly limited by the sensitivity of the 
detectors at the corresponding BH frequencies,
that occur at $f\gtrsim 6~{\rm kHz}$
for typical BNS systems.

Figure~\ref{fig:pms} (bottom panel)
shows the posterior distributions
for the frequency drift $\adrift$
as functions of the injected SNR.
Differently from $\tc$,
this term appears to be better constrained
for long-lived equal-mass binaries.
This is due to the nature of these PM GW 
transients. In fact, their spectra
are characterized by the loudest PM peaks,
permitting an improved identification 
of the peak properties compared to the 
short-lived remnant cases.
Also $\adrift$ is affected by noise fluctuations,
affecting primarily the widths of the recovered
posteriors. However, for SNR $\gtrsim 10$,
systematic errors become more relevant.
A particular case is LS220 binary, 
for which the recovered posterior overestimates
the nominal injected value.
However, as previously discussed, 
this case has non-monotonic 
frequency evolution. Thus, {\model} 
is recovering
the initial slope of the LS220 transient,
corresponding to the loudest contribution.
For unequal-mass binary, $\adrift$
is poorly constrained due to the faintness
of the late-time portion of the GW transient.

\section{EOS softening tests}
\label{sec:softness}

\begin{figure}[t]
	\centering 
	\includegraphics[width=0.49\textwidth]{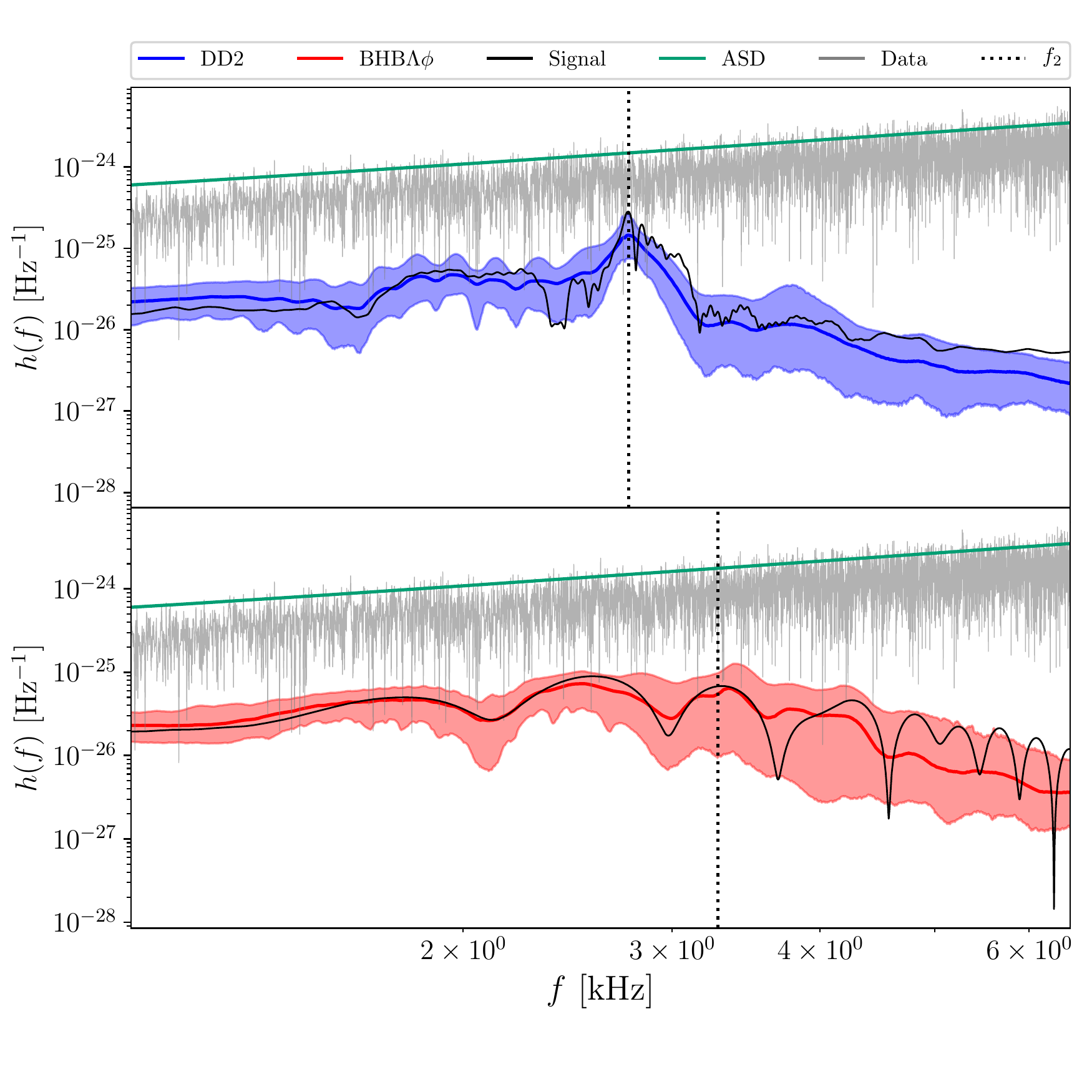}
	\caption{Posterior distributions 
					of the recovered spectra
					 for the DD2 (blue)
					 and the BHB$\Lambda\phi$ (red)
					 binaries at PM SNR 8.
			 		The solid colored lines report 
			 		the median spectrum and the 
			 		 contours show the 90\% credible regions. 
			 		The black solid lines correspond to the injected signals 
			 		and black dotted lines correspond
			 		to the $f_2$ peak.
			 		The green lines show the amplitude spectra
			 		density (ASD).
			 		The gray lines show the artificial data. 
	}
	\label{fig:soft_snr8}
\end{figure}

\begin{figure}[t]
	\centering 
	\includegraphics[width=0.49\textwidth]{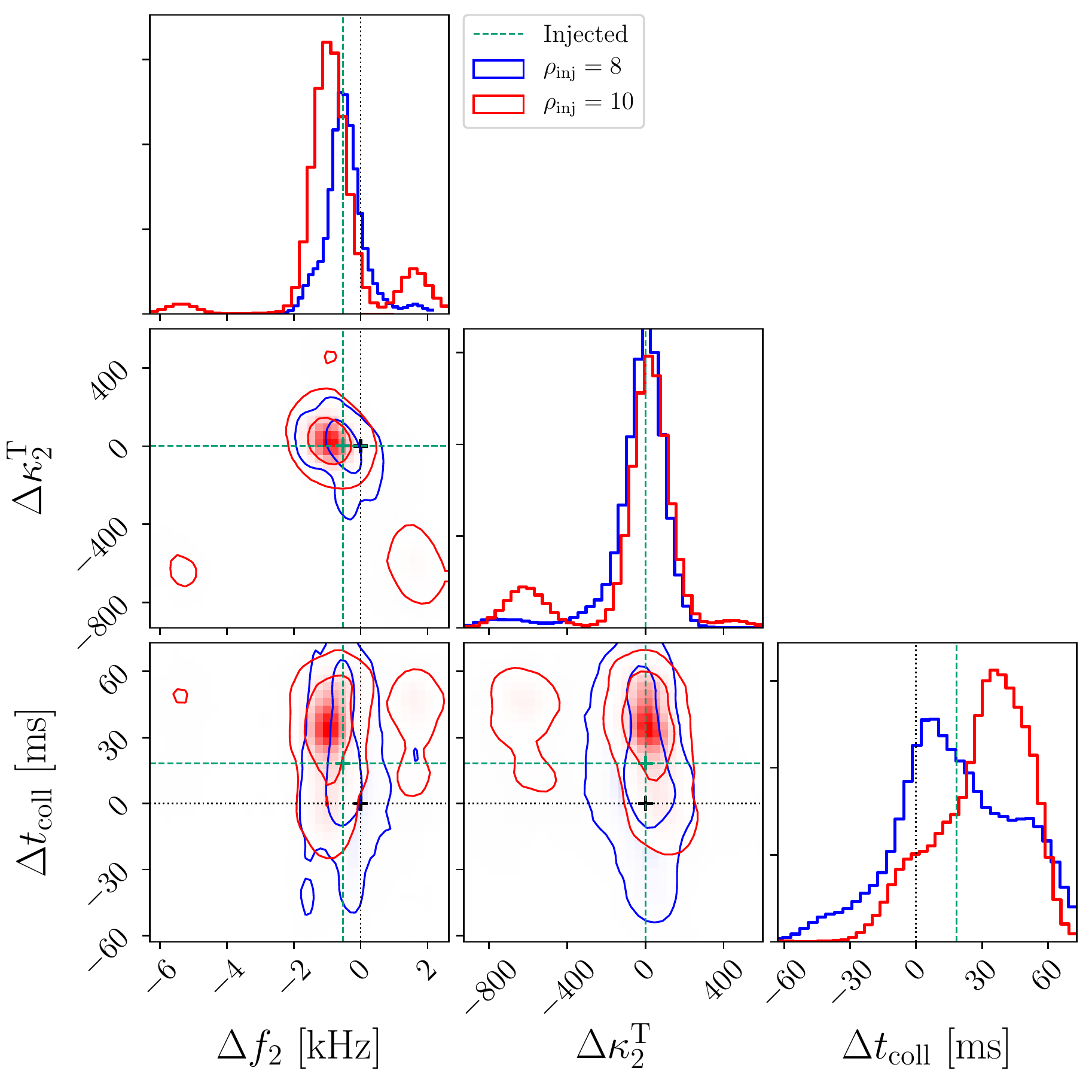}
	\caption{Posterior distribution
		for the deviations $\{\Delta f_2, \Delta\kt , \Delta\tc\}$ between the DD2 and the BHB$\Lambda\phi$
		analyses
		for PM SNR 8 (blue) and 10 (red).
		The contours correspond
		to the 50\% annd 90\% credibility 
		regions.
		The injected deviations are reported 
		with green dashed lines and 
		the zero hypothesis is 
		reported with black dotted lines.
		A deviation is defined 
		as $\Delta Q = Q^{\rm DD2}-Q^{{\rm BHB}\Lambda\phi}$ for each 
		quantity $Q$. 
	    }
	\label{fig:soft_deviation}
\end{figure}

In this Section, we study the possibility of inferring an EOS
softnening from a detection of PM signal. Specifically, we repeat the analysis presented in Sec.~{V D} of
\cite{Breschi:2019srl} with the updated model {\model} and employing ET detector. 
The injections
are performed within the same framework discussed in
Sec.~\ref{sec:framework} with 
PM SNR $\rho_{\rm inj} = \{8,10\}$ 
using a
single noise realization. We perform PE on the PM transients solely; the latter are  
computed from NR simulations of 
$1.50{+}1.50~\Msun$ BNS systems described by DD2 and BHB$\Lambda\phi$
EOSs~\cite{Radice:2016rys}. The two EOSs are identical except for the
inclusion of hyperonic degrees of freedom at high densities in
BHB$\Lambda\phi$~\cite{Banik:2014qja}. 
This inclusion introduces a softening 
at the extreme densities reached in the remnant and deviates from DD2 at 
typical 
densities $\rho\simeq 2\rho_{\rm sat}$,
where $\rho_{\rm sat}=2.7{\times }10^{14}~{\rm g}~{\rm cm}^{-3}$ is the nuclear saturation density. 
The main question we address is whether an analysis of PM signal can
inform us on the EOS softening effects at extreme densities, and
  what are the most relevant quantities that indicate the softening.

Previous studies performed with {\oldmodel} showed that the $f_2$
posterior of the BHB$\Lambda\phi$ strongly differs from the DD2 case
\cite{Breschi:2019srl}. We remark that the DD2 $1.50{+}1.50~\Msun$
binary has $f_2\simeq 2.76~{\rm kHz}$; while, the respective
BHB$\Lambda\phi$ case has $f_2\simeq 3.29~{\rm kHz}$~\footnote{
  As discussed in Paper~I, the precise identification of $f_2$ is
  challenging for this binary due to the short PM signal. Here, we assume
  the second peak of the spectrum is $f_2$.}. 
The difference
between the two NR values is ${\sim} 500~{\rm Hz}$, which corresponds
to ${\sim }15\%$. The BHB$\Lambda\phi$ data deviates of ${\sim}3{-}\sigma$
from  the prediction of the EOS-insensitive relation presented in {\paperI} 
($f_2^{\rm fit}=2.88~{\rm kHz}$), encoding a more compact remnant than the DD2 case. 
The two binaries have also different times of BH collapse: the DD2 case collapses at late times, i.e. $\tc\simeq 21~{\rm ms}$; while, the BHB$\Lambda\phi$ remnant collapses into BH shortly after merger with $\tc\simeq 2.6~{\rm ms}$.
The BHB$\Lambda\phi$ $1.50{+}1.50~\Msun$ binary shows the largest deviations from the quasiuniversal fits used in {\model}.

The spectra recovered in the 
new analyses performed with {\model} 
at PM SNR 8 are shown 
in Fig.~\ref{fig:soft_snr8}.
The recovered SNRs are consistent with 
the injected values, 
recovering
$\rho_{\rm rec}={7.5}^{+1.0}_{-1.4}$
and a BF $\log\B=6.3\pm0.2$
for the DD2 case;
while, for the BHB$\Lambda\phi$,
we get
$\rho_{\rm rec}={7.4}^{+1.2}_{-1.2}$
and $\log\B=6.5\pm0.2$.
The posterior distributions 
for the intrinsic binary parameters 
do not show significant 
differences between the two cases.
This is key in view of 
coherent and informative consistency tests~
\citep[e.g.][]{Breschi:2019srl,
	Wijngaarden:2022sah}.
We recover $M\simeq 4.3~\Msun$
with an error of ${\sim}1.5~\Msun$.
The mass ratios are constrained around the equal-mass
case with errors of ${\sim}0.25$.
The tidal polarizability $\kt$
underestimates the injected values
with medians $\kt \simeq 30$ and errors
${\sim}150$.
The spins posteriors are dominated by the priors and not informative.
	
At SNR 8,
the $f_2$ posteriors show different
median values corresponding to 
$f_2={2.73}^{+0.05}_{-0.06}~{\rm kHz}$
for DD2 and
$f_2={3.2}^{+0.9}_{-1.0}~{\rm kHz}$
for BHB$\Lambda\phi$.
The difference
between the recovered $f_2$
exceed the $1{-}\sigma$ credibility level
of the $f_2$ EOS-insensitive relation,
highlighting a breaking of quasiuniversality
within that level.
This deviation is partially encoded 
in the related recalibration parameters
that recovered opposite values, i.e.
$\delta_{f_2}={-1}^{+7}_{-4}~\%$ 
for DD2 and
$\delta_{f_2}={2}^{+6}_{-9}~\%$ 
for BHB$\Lambda\phi$.
In this context,
the inclusion of the recalibration coefficients
is crucial since they 
allow the observed PM peak to deviate
from the prediction of the quasiuniversal 
relation for a common combination of 
intrinsic binary parameters.
However, the short duration of the BHB$\Lambda\phi$
transient and the strong modulations
can introduce biases and 
multimodalities
in the estimates of the 
associated peak frequency, 
as also shown in the analysis of the BLQ binary 
in Sec.~\ref{sec:nr}.
	
Another quantity that encodes the softening of the 
EOS at high densities is the time of BH collapse 
$\tc$. 
A softer EOS allows the NS remnant 
to reach higher densities, yielding to an earlier BH
collapse for comparable masses~\cite{Radice:2016rys,
Prakash:2021wpz,
Fujimoto:2022xhv}.
The recovered posteriors at SNR 8 for $\tc$
give $\tc={28}^{+36}_{-24}~{\rm ms}$
for DD2 and 
$\tc={6}^{+41}_{-4}~{\rm ms}$
for BHB$\Lambda\phi$,
consistently with the injected values.
Even if this term can be strongly affected 
by noise fluctuations at low SNRs, as discussed in Sec.~\ref{sec:pmparams},
the estimate of $\tc$ appears to be 
less biased and more 
conservative compared to the $f_2$ one.
Moreover,
the $\tc$ posterior is expected to 
converge to the true value for increasing SNR
for short-lived remnant, as shown by the BLQ
binary in Sec.~\ref{sec:nr}.
Thus, 
the measurement of $\tc$ is expected
to be a robust probe of softening effects 
in the NS EOS for PM SNR $\gtrsim 9$.

In order to 
validate the robustness of the 
inference, 
we repeat the analysis 
injecting the signals at PM SNR 10.
The recovered SNRs are
$\rho_{\rm rec}={7.8}^{+1.2}_{-1.2}$
and $\log\B=7.0\pm0.2$
for the DD2 case
and 
$\rho_{\rm rec}={7.6}^{+1.8}_{-1.9}$
and $\log\B=9.6\pm0.2$
for the BHB$\Lambda\phi$ case.
The dominant systematic appears 
in the $f_2$ posterior distribution for BHB$\Lambda\phi$.
This posteriors shows pronounced
multimodalities
due to the contribution
of the subdominant coupled frequencies $f_{2\pm0}$,
as discussed previously.
The median values and the $90\%$
credibility intervals correspond to 
$f_2={2.75}^{+0.03}_{-0.03}~{\rm kHz}$
for the DD2 binary and 
$f_2={3.7}^{+0.3}_{-2.7}~{\rm kHz}$
for the BHB$\Lambda\phi$ binary. 
The $\tc$ posterior
distributions show convergent trends 
toward the injected values for both cases,
recovering 
$\tc={54}^{+5}_{-41}~{\rm ms}$
for the DD2 binary and 
$\tc={16}^{+11}_{-13}~{\rm ms}$
for the BHB$\Lambda\phi$ binary.
From these results,
it is possible to understand that 
the short-lived binaries 
(such as the BHB$\Lambda\phi$ case)
show more ambiguous $f_2$ posteriors
and more informative $\tc$ posteriors
for increasing SNRs.
On the other hand,
for the 
long-lived binaries (such as the DD2 case),
the $f_2$ posterior is typically unbiased 
and the $\tc$ posterior is more affected 
by noise fluctuations,
but showing convergence toward the injected value
and typically 
including the latter within the 90\% credibility 
intervals.

Aiming to highlight the differences
between the DD2 and BHB$\Lambda\phi$ 
posterior distributions,
we introduce the deviation 
$\Delta Q = Q^{\rm DD2} - Q^{{\rm BHB}\Lambda\phi}$,
where $Q$ is an arbitrary PM quantity.
Figure~\ref{fig:soft_deviation}
shows the posterior distributions
for $\{\Delta f_2, \Delta\kt , \Delta\tc\}$
for the analyses with SNR 8 and 10.
In a realistic scenario,
such computation is unlikely since it would 
require the observation of two identical BNS
mergers with different EOS.
However, 
in order to understand the performances
of the model,
this representation clearly evinces 
the differences between the 
recovered posteriors.
The $\Delta f_2$ posterior reports
mild deviations for SNR 8, 
i.e. $\Delta f_2 = {-0.5}^{+1.0}_{-0.9}~{\rm kHz}$; while, for SNR 10,
the average deviation increases,
i.e. $\Delta f_2 = {-0.9}^{+2.6}_{-0.9}~{\rm kHz}$,
but the posterior introduces ambiguities
due to multimodalities.
The $\Delta\kt$ posteriors confidently 
include the zero hypothesis within the 
support
and it shows small
multimodalities due to the correlation
with the PM frequency $f_2$
at SNR 10.
On the other hand,
the $\Delta\tc$ posteriors
show unambiguous 
convergence toward a non-zero
value for increasing SNR.
The median values and the 90\% credibility 
intervals are 
$\Delta \tc = {16}^{+46}_{-44}~{\rm ms}$
at SNR 8
and 
$\Delta \tc = {35}^{+23}_{-42}~{\rm ms}$
at SNR 10.

\section{Conclusions}
\label{sec:conclusion}

We presented PE studies on PM GW transients
solely from BNS merger remnants with the novel
template model {\model}
focusing on next-generation detector ET~\cite{Hild:2010id,Hild:2011np}.
Performing a survey of $245$ PE studies
with seven different artificial templates 
and five noise realizations,
we demonstrated that ET can detect PM signals from a threshold
SNR ${\sim}7$ using {\model} template and without any pre-merger information.
These SNRs correspond
to (optimally-oriented) binaries
located at luminosity distances of ${\sim}100~{\rm Mpc}$,
that are values consistent with recent observations of BNS mergers~\cite{Abbott:2020uma}.
Employing estimates of BNS merger rates~\cite{LIGOScientific:2021psn}, 
we compute an upper limit of ${\sim}4$ 
observable BNS 
mergers per year in a 
spherical volume of radius $100~{\rm Mpc}$.

The inference with the novel {\model} improves over the previous {\oldmodel}~\cite{Breschi:2019srl,Breschi:2021xrx}
due to the improved calibration of EOS-sensitive relations performed
in {\paperI}. The inference results are consistent with the
expectations from the faithfulness studies
reported in {\paperI}.
We have shown that {\model} can provide 
constraints on several characteristic properties
of the observed PM signals. 
The uncertainties on the $f_2$ posteriors 
are $O(1~{\rm kHz})$ at PM SNR 7
and $O(100~{\rm Hz})$ at PM SNR 10.
However, systematic errors arise 
for short-lived and large-mass-ratio 
remnants for SNR $\gtrsim 8$
primarily related 
to an inaccurate identification of the
characteristic PM peak.
The time of 
BH collapse $\tc$ can be estimated 
with an accuracy of $O(20~{\rm ms})$
at PM SNR 10.
The inference of this term
is significantly affected by the employed
noise realization,
since the late-time tail of the PM signal is
masked by noise fluctuations.
Moreover, {\model} can provide 
measurements of the subdominant coupling frequency $f_0$,
of the frequency drift $\adrift$ 
and on the intrinsic BNS properties;
however, the latter appears to be generally poorly 
constrained compared to a corresponding pre-merger
analysis~\cite[e.g.][]{Breschi:2019srl,Wijngaarden:2022sah}.

Our injection survey with {\model} indicates that 
the largest biases and systematic errors 
arise for 
very-short-lived remnants
and tidally-disruptive mergers.
These errors are primarily related to
misinterpretation of the 
subdominant coupling frequencies
or 
to an insufficient GW power in the PM segment.
The systematic biases 
can be mitigated 
improving the 
model morphology
 in the large-mass-ratio
regimes,
when more NR simulations become
available,
and including the wavelet 
corresponding to the BH ringdown
(see {\paperI}).
However, a full description of the 
PM collapse dynamics in BNS mergers
will require a more accurate 
characterization and modeling,
in particular for 
large-mass-ratio mergers,
i.e. $q>1.5$,
and massive remnants, 
i.e. $M\gtrsim 2.8~\Msun$.
Nevertheless, the direct observation 
of the BH collapse and the subsequent ringdown 
radiation are considerably limited by the detector sensitivities 
in these extreme region of the Fourier spectrum, i.e. $f\gtrsim 6~{\rm kHz}$.
Thus, a neat observation
of a BH ringdown after a BNS merger 
is challenging.

PM signals can, in principle, inform us on the appearance of non-nucleonic 
degrees of freedom at extreme matter densities. A possible imprint of a EOS softening in the signal is the breaking of the quasiuniversal relations that characterize the spectral features. 
For example, several NR simulations indicate that deviations
of the orders of ${\sim}500~{\rm Hz}$ 
in the $f_2$ peak
and of ${\sim}20~{\rm ms}$ in the time
of BH collapse are possible due EOS softening, \citep[e.g.][]{Sekiguchi:2011mc,Radice:2016rys,Bauswein:2018bma,Prakash:2021wpz,Fujimoto:2022xhv}.
We considered a short-lived remnant as a case study 
  and demonstrate that ${\sim}3{-}\sigma$ violations of EOS-insensitive relations  
are potentially observable at PM SNR 8.
However, the 
short duration of the PM transient might lead to 
biases in the inference of $f_2$
for SNR $\gtrsim 9$.
On the other hand, the 
inference of the time of collapse $\tc$
delivers more robust measurements for 
increasing SNRs for short-lived remnant. 
We stress that the breaking of a quasiuniversal relation within a given confidence level does not necessarily imply the presence of an EOS softening~\footnote{A counterexample is given in the conclusion of {\paperI}}.
 Instead, more generally, such a breakdown should be interpreted as the invalidation of a particular 
 	empirical relation 
 	due to physical effects 
 	that not captured by the 
 	constructed fit.
Similarly, a EOS softening might break a quasiuniversal relation only in a marginally significant way, 
as in our case study,
or the remnant's densities might be such that these effects are not evident,
e.g. small mass binaries $M\ll 3~\Msun$~\cite{Radice:2016rys,Prakash:2021wpz}.
In the future, it would be interesting to verify the 
performances of {\model} against NR templates
computed with more extreme EOS that show larger deviations from 
the quasiuniversal trends~\citep[e.g.][]{Bauswein:2018bma,
	Wijngaarden:2022sah}.
In general, more accurate studies 
of BNS merger with non-nucleonic EOS,
driven by high-precision NR simulations~\citep[e.g.][]{Radice:2016rys,
	Prakash:2021wpz,
Fujimoto:2022xhv},
are essential in order to 
comprehensively characterize to what extend the breaking of EOS-insensitive relations can probe EOS softening.

Finally, our results should be revisited with 
a full inspiral-merger-postmerger analysis, 
similarly to what
  presented in Ref.~\cite{Breschi:2019srl,Wijngaarden:2022sah}.
These studies will be reported in the third paper of this series.
The pre-merger information
is expected to provide 
narrower constraints
on the intrinsic binary parameters
and to reduce the related biases~\cite{Breschi:2019srl,Wijngaarden:2022sah}.
This will have a significant effects on the 
overall inference, e.g.
improving the identification of the correlations 
between the intrinsic and the PM parameters.
Moreover, 
the recalibration parameters are expected to 
have a more relevant effect on the recovered posteriors,
highlighting more noticeable deviations from quasiuniversal behavior.
The pre-merger information will also contribute 
to the inference high-density
EOS properties~\cite{Breschi:2021xrx}
and to the identification
of softening effects, allowing coherent
pre-and-post-merger 
consistency tests~\cite{Breschi:2019srl}.


\section*{Acknowledgments}

MB, SB and GC acknowledge 
support by the EU H2020 under ERC Starting
Grant, no.~BinGraSp-714626.
MB and RG acknowledges support from the Deutsche Forschungsgemeinschaft
(DFG) under Grant No. 406116891 within the Research Training Group
RTG 2522/1.
SB acknowledges support from the Deutsche Forschungsgemeinschaft, DFG,
project MEMI number BE 6301/2-1.
The computational experiments were performed on {\scshape ARA},
a resource of Friedrich-Schiller-Universt\"at Jena supported in part by DFG grants INST 275/334-1 FUGG, INST 275/363-1 FUGG and EU H2020 BinGraSp-714626.
Postprocessing was performed on the {\scshape Tullio} sever at INFN Turin.

The waveform model developed in this work, {\model}, 
is implemented in {\scshape bajes} and the software is publicly available at:

\url{https://github.com/matteobreschi/bajes}


\appendix


\begin{thebibliography}{43}%
\makeatletter
\providecommand \@ifxundefined [1]{%
 \@ifx{#1\undefined}
}%
\providecommand \@ifnum [1]{%
 \ifnum #1\expandafter \@firstoftwo
 \else \expandafter \@secondoftwo
 \fi
}%
\providecommand \@ifx [1]{%
 \ifx #1\expandafter \@firstoftwo
 \else \expandafter \@secondoftwo
 \fi
}%
\providecommand \natexlab [1]{#1}%
\providecommand \enquote  [1]{``#1''}%
\providecommand \bibnamefont  [1]{#1}%
\providecommand \bibfnamefont [1]{#1}%
\providecommand \citenamefont [1]{#1}%
\providecommand \href@noop [0]{\@secondoftwo}%
\providecommand \href [0]{\begingroup \@sanitize@url \@href}%
\providecommand \@href[1]{\@@startlink{#1}\@@href}%
\providecommand \@@href[1]{\endgroup#1\@@endlink}%
\providecommand \@sanitize@url [0]{\catcode `\\12\catcode `\$12\catcode
  `\&12\catcode `\#12\catcode `\^12\catcode `\_12\catcode `\%12\relax}%
\providecommand \@@startlink[1]{}%
\providecommand \@@endlink[0]{}%
\providecommand \url  [0]{\begingroup\@sanitize@url \@url }%
\providecommand \@url [1]{\endgroup\@href {#1}{\urlprefix }}%
\providecommand \urlprefix  [0]{URL }%
\providecommand \Eprint [0]{\href }%
\providecommand \doibase [0]{http://dx.doi.org/}%
\providecommand \selectlanguage [0]{\@gobble}%
\providecommand \bibinfo  [0]{\@secondoftwo}%
\providecommand \bibfield  [0]{\@secondoftwo}%
\providecommand \translation [1]{[#1]}%
\providecommand \BibitemOpen [0]{}%
\providecommand \bibitemStop [0]{}%
\providecommand \bibitemNoStop [0]{.\EOS\space}%
\providecommand \EOS [0]{\spacefactor3000\relax}%
\providecommand \BibitemShut  [1]{\csname bibitem#1\endcsname}%
\let\auto@bib@innerbib\@empty
\bibitem [{\citenamefont {Hild}\ \emph {et~al.}(2008)\citenamefont {Hild},
  \citenamefont {Chelkowski},\ and\ \citenamefont {Freise}}]{Hild:2008ng}%
  \BibitemOpen
  \bibfield  {author} {\bibinfo {author} {\bibfnamefont {S.}~\bibnamefont
  {Hild}}, \bibinfo {author} {\bibfnamefont {S.}~\bibnamefont {Chelkowski}}, \
  and\ \bibinfo {author} {\bibfnamefont {A.}~\bibnamefont {Freise}},\
  }\href@noop {} {\  (\bibinfo {year} {2008})},\ \Eprint
  {http://arxiv.org/abs/0810.0604} {arXiv:0810.0604 [gr-qc]} \BibitemShut
  {NoStop}%
\bibitem [{\citenamefont {Radice}\ \emph {et~al.}(2020)\citenamefont {Radice},
  \citenamefont {Bernuzzi},\ and\ \citenamefont {Perego}}]{Radice:2020ddv}%
  \BibitemOpen
  \bibfield  {author} {\bibinfo {author} {\bibfnamefont {D.}~\bibnamefont
  {Radice}}, \bibinfo {author} {\bibfnamefont {S.}~\bibnamefont {Bernuzzi}}, \
  and\ \bibinfo {author} {\bibfnamefont {A.}~\bibnamefont {Perego}},\ }\href
  {\doibase 10.1146/annurev-nucl-013120-114541} {\bibfield  {journal} {\bibinfo
   {journal} {Ann. Rev. Nucl. Part. Sci.}\ }\textbf {\bibinfo {volume} {70}}
  (\bibinfo {year} {2020}),\ 10.1146/annurev-nucl-013120-114541},\ \Eprint
  {http://arxiv.org/abs/2002.03863} {arXiv:2002.03863 [astro-ph.HE]}
  \BibitemShut {NoStop}%
\bibitem [{\citenamefont {Bernuzzi}(2020)}]{Bernuzzi:2020tgt}%
  \BibitemOpen
  \bibfield  {author} {\bibinfo {author} {\bibfnamefont {S.}~\bibnamefont
  {Bernuzzi}},\ }\href {\doibase 10.1007/s10714-020-02752-5} {\bibfield
  {journal} {\bibinfo  {journal} {Gen. Rel. Grav.}\ }\textbf {\bibinfo {volume}
  {52}},\ \bibinfo {pages} {108} (\bibinfo {year} {2020})},\ \Eprint
  {http://arxiv.org/abs/2004.06419} {arXiv:2004.06419 [astro-ph.HE]}
  \BibitemShut {NoStop}%
\bibitem [{\citenamefont {Chatziioannou}\ \emph {et~al.}(2017)\citenamefont
  {Chatziioannou}, \citenamefont {Clark}, \citenamefont {Bauswein},
  \citenamefont {Millhouse}, \citenamefont {Littenberg},\ and\ \citenamefont
  {Cornish}}]{Chatziioannou:2017ixj}%
  \BibitemOpen
  \bibfield  {author} {\bibinfo {author} {\bibfnamefont {K.}~\bibnamefont
  {Chatziioannou}}, \bibinfo {author} {\bibfnamefont {J.~A.}\ \bibnamefont
  {Clark}}, \bibinfo {author} {\bibfnamefont {A.}~\bibnamefont {Bauswein}},
  \bibinfo {author} {\bibfnamefont {M.}~\bibnamefont {Millhouse}}, \bibinfo
  {author} {\bibfnamefont {T.~B.}\ \bibnamefont {Littenberg}}, \ and\ \bibinfo
  {author} {\bibfnamefont {N.}~\bibnamefont {Cornish}},\ }\href {\doibase
  10.1103/PhysRevD.96.124035} {\bibfield  {journal} {\bibinfo  {journal} {Phys.
  Rev.}\ }\textbf {\bibinfo {volume} {D96}},\ \bibinfo {pages} {124035}
  (\bibinfo {year} {2017})},\ \Eprint {http://arxiv.org/abs/1711.00040}
  {arXiv:1711.00040 [gr-qc]} \BibitemShut {NoStop}%
\bibitem [{\citenamefont {Tsang}\ \emph {et~al.}(2019)\citenamefont {Tsang},
  \citenamefont {Dietrich},\ and\ \citenamefont {Van
  Den~Broeck}}]{Tsang:2019esi}%
  \BibitemOpen
  \bibfield  {author} {\bibinfo {author} {\bibfnamefont {K.~W.}\ \bibnamefont
  {Tsang}}, \bibinfo {author} {\bibfnamefont {T.}~\bibnamefont {Dietrich}}, \
  and\ \bibinfo {author} {\bibfnamefont {C.}~\bibnamefont {Van Den~Broeck}},\
  }\href {\doibase 10.1103/PhysRevD.100.044047} {\bibfield  {journal} {\bibinfo
   {journal} {Phys. Rev.}\ }\textbf {\bibinfo {volume} {D100}},\ \bibinfo
  {pages} {044047} (\bibinfo {year} {2019})},\ \Eprint
  {http://arxiv.org/abs/1907.02424} {arXiv:1907.02424 [gr-qc]} \BibitemShut
  {NoStop}%
\bibitem [{\citenamefont {Breschi}\ \emph {et~al.}(2019)\citenamefont
  {Breschi}, \citenamefont {Bernuzzi}, \citenamefont {Zappa}, \citenamefont
  {Agathos}, \citenamefont {Perego}, \citenamefont {Radice},\ and\
  \citenamefont {Nagar}}]{Breschi:2019srl}%
  \BibitemOpen
  \bibfield  {author} {\bibinfo {author} {\bibfnamefont {M.}~\bibnamefont
  {Breschi}}, \bibinfo {author} {\bibfnamefont {S.}~\bibnamefont {Bernuzzi}},
  \bibinfo {author} {\bibfnamefont {F.}~\bibnamefont {Zappa}}, \bibinfo
  {author} {\bibfnamefont {M.}~\bibnamefont {Agathos}}, \bibinfo {author}
  {\bibfnamefont {A.}~\bibnamefont {Perego}}, \bibinfo {author} {\bibfnamefont
  {D.}~\bibnamefont {Radice}}, \ and\ \bibinfo {author} {\bibfnamefont
  {A.}~\bibnamefont {Nagar}},\ }\href {\doibase 10.1103/PhysRevD.100.104029}
  {\bibfield  {journal} {\bibinfo  {journal} {Phys. Rev.}\ }\textbf {\bibinfo
  {volume} {D100}},\ \bibinfo {pages} {104029} (\bibinfo {year} {2019})},\
  \Eprint {http://arxiv.org/abs/1908.11418} {arXiv:1908.11418 [gr-qc]}
  \BibitemShut {NoStop}%
\bibitem [{\citenamefont {Easter}\ \emph {et~al.}(2020)\citenamefont {Easter},
  \citenamefont {Ghonge}, \citenamefont {Lasky}, \citenamefont {Casey},
  \citenamefont {Clark}, \citenamefont {Vivanco},\ and\ \citenamefont
  {Chatziioannou}}]{Easter:2020ifj}%
  \BibitemOpen
  \bibfield  {author} {\bibinfo {author} {\bibfnamefont {P.~J.}\ \bibnamefont
  {Easter}}, \bibinfo {author} {\bibfnamefont {S.}~\bibnamefont {Ghonge}},
  \bibinfo {author} {\bibfnamefont {P.~D.}\ \bibnamefont {Lasky}}, \bibinfo
  {author} {\bibfnamefont {A.~R.}\ \bibnamefont {Casey}}, \bibinfo {author}
  {\bibfnamefont {J.~A.}\ \bibnamefont {Clark}}, \bibinfo {author}
  {\bibfnamefont {F.~H.}\ \bibnamefont {Vivanco}}, \ and\ \bibinfo {author}
  {\bibfnamefont {K.}~\bibnamefont {Chatziioannou}},\ }\href {\doibase
  10.1103/PhysRevD.102.043011} {\bibfield  {journal} {\bibinfo  {journal}
  {Phys. Rev. D}\ }\textbf {\bibinfo {volume} {102}},\ \bibinfo {pages}
  {043011} (\bibinfo {year} {2020})},\ \Eprint
  {http://arxiv.org/abs/2006.04396} {arXiv:2006.04396 [astro-ph.HE]}
  \BibitemShut {NoStop}%
\bibitem [{\citenamefont {Breschi}\ \emph
  {et~al.}(2022{\natexlab{a}})\citenamefont {Breschi}, \citenamefont
  {Bernuzzi}, \citenamefont {Godzieba}, \citenamefont {Perego},\ and\
  \citenamefont {Radice}}]{Breschi:2021xrx}%
  \BibitemOpen
  \bibfield  {author} {\bibinfo {author} {\bibfnamefont {M.}~\bibnamefont
  {Breschi}}, \bibinfo {author} {\bibfnamefont {S.}~\bibnamefont {Bernuzzi}},
  \bibinfo {author} {\bibfnamefont {D.}~\bibnamefont {Godzieba}}, \bibinfo
  {author} {\bibfnamefont {A.}~\bibnamefont {Perego}}, \ and\ \bibinfo {author}
  {\bibfnamefont {D.}~\bibnamefont {Radice}},\ }\href {\doibase
  10.1103/PhysRevLett.128.161102} {\bibfield  {journal} {\bibinfo  {journal}
  {Phys. Rev. Lett.}\ }\textbf {\bibinfo {volume} {128}},\ \bibinfo {pages}
  {161102} (\bibinfo {year} {2022}{\natexlab{a}})},\ \Eprint
  {http://arxiv.org/abs/2110.06957} {arXiv:2110.06957 [gr-qc]} \BibitemShut
  {NoStop}%
\bibitem [{\citenamefont {Wijngaarden}\ \emph {et~al.}(2022)\citenamefont
  {Wijngaarden}, \citenamefont {Chatziioannou}, \citenamefont {Bauswein},
  \citenamefont {Clark},\ and\ \citenamefont {Cornish}}]{Wijngaarden:2022sah}%
  \BibitemOpen
  \bibfield  {author} {\bibinfo {author} {\bibfnamefont {M.}~\bibnamefont
  {Wijngaarden}}, \bibinfo {author} {\bibfnamefont {K.}~\bibnamefont
  {Chatziioannou}}, \bibinfo {author} {\bibfnamefont {A.}~\bibnamefont
  {Bauswein}}, \bibinfo {author} {\bibfnamefont {J.~A.}\ \bibnamefont {Clark}},
  \ and\ \bibinfo {author} {\bibfnamefont {N.~J.}\ \bibnamefont {Cornish}},\
  }\href@noop {} {\  (\bibinfo {year} {2022})},\ \Eprint
  {http://arxiv.org/abs/2202.09382} {arXiv:2202.09382 [gr-qc]} \BibitemShut
  {NoStop}%
\bibitem [{\citenamefont {Veitch}\ \emph {et~al.}(2015)\citenamefont {Veitch}
  \emph {et~al.}}]{Veitch:2014wba}%
  \BibitemOpen
  \bibfield  {author} {\bibinfo {author} {\bibfnamefont {J.}~\bibnamefont
  {Veitch}} \emph {et~al.},\ }\href {\doibase 10.1103/PhysRevD.91.042003}
  {\bibfield  {journal} {\bibinfo  {journal} {Phys. Rev.}\ }\textbf {\bibinfo
  {volume} {D91}},\ \bibinfo {pages} {042003} (\bibinfo {year} {2015})},\
  \Eprint {http://arxiv.org/abs/1409.7215} {arXiv:1409.7215 [gr-qc]}
  \BibitemShut {NoStop}%
\bibitem [{\citenamefont {Lange}\ \emph {et~al.}(2018)\citenamefont {Lange},
  \citenamefont {O'Shaughnessy},\ and\ \citenamefont {Rizzo}}]{Lange:2018pyp}%
  \BibitemOpen
  \bibfield  {author} {\bibinfo {author} {\bibfnamefont {J.}~\bibnamefont
  {Lange}}, \bibinfo {author} {\bibfnamefont {R.}~\bibnamefont
  {O'Shaughnessy}}, \ and\ \bibinfo {author} {\bibfnamefont {M.}~\bibnamefont
  {Rizzo}},\ }\href@noop {} {\  (\bibinfo {year} {2018})},\ \Eprint
  {http://arxiv.org/abs/1805.10457} {arXiv:1805.10457 [gr-qc]} \BibitemShut
  {NoStop}%
\bibitem [{\citenamefont {Breschi}\ \emph {et~al.}(2021)\citenamefont
  {Breschi}, \citenamefont {Gamba},\ and\ \citenamefont
  {Bernuzzi}}]{Breschi:2021wzr}%
  \BibitemOpen
  \bibfield  {author} {\bibinfo {author} {\bibfnamefont {M.}~\bibnamefont
  {Breschi}}, \bibinfo {author} {\bibfnamefont {R.}~\bibnamefont {Gamba}}, \
  and\ \bibinfo {author} {\bibfnamefont {S.}~\bibnamefont {Bernuzzi}},\ }\href
  {\doibase 10.1103/PhysRevD.104.042001} {\bibfield  {journal} {\bibinfo
  {journal} {Phys. Rev. D}\ }\textbf {\bibinfo {volume} {104}},\ \bibinfo
  {pages} {042001} (\bibinfo {year} {2021})},\ \Eprint
  {http://arxiv.org/abs/2102.00017} {arXiv:2102.00017 [gr-qc]} \BibitemShut
  {NoStop}%
\bibitem [{\citenamefont {Breschi}\ \emph
  {et~al.}(2022{\natexlab{b}})\citenamefont {Breschi} \emph
  {et~al.}}]{Breschi:2022}%
  \BibitemOpen
  \bibfield  {author} {\bibinfo {author} {\bibfnamefont {M.}~\bibnamefont
  {Breschi}} \emph {et~al.},\ }\href@noop {} {\bibfield  {journal} {\bibinfo
  {journal} {{\it (in preparation)}}\ } (\bibinfo {year}
  {2022}{\natexlab{b}})}\BibitemShut {NoStop}%
\bibitem [{\citenamefont {Hild}\ \emph {et~al.}(2011)\citenamefont {Hild} \emph
  {et~al.}}]{Hild:2010id}%
  \BibitemOpen
  \bibfield  {author} {\bibinfo {author} {\bibfnamefont {S.}~\bibnamefont
  {Hild}} \emph {et~al.},\ }\href {\doibase 10.1088/0264-9381/28/9/094013}
  {\bibfield  {journal} {\bibinfo  {journal} {Class. Quant. Grav.}\ }\textbf
  {\bibinfo {volume} {28}},\ \bibinfo {pages} {094013} (\bibinfo {year}
  {2011})},\ \Eprint {http://arxiv.org/abs/1012.0908} {arXiv:1012.0908 [gr-qc]}
  \BibitemShut {NoStop}%
\bibitem [{\citenamefont {Hild}(2012)}]{Hild:2011np}%
  \BibitemOpen
  \bibfield  {author} {\bibinfo {author} {\bibfnamefont {S.}~\bibnamefont
  {Hild}},\ }\href {\doibase 10.1088/0264-9381/29/12/124006} {\bibfield
  {journal} {\bibinfo  {journal} {Class.Quant.Grav.}\ }\textbf {\bibinfo
  {volume} {29}},\ \bibinfo {pages} {124006} (\bibinfo {year} {2012})},\
  \Eprint {http://arxiv.org/abs/1111.6277} {arXiv:1111.6277 [gr-qc]}
  \BibitemShut {NoStop}%
\bibitem [{\citenamefont {Punturo}\ \emph
  {et~al.}(2010{\natexlab{a}})\citenamefont {Punturo}, \citenamefont
  {Abernathy}, \citenamefont {Acernese}, \citenamefont {Allen}, \citenamefont
  {Andersson} \emph {et~al.}}]{Punturo:2010zz}%
  \BibitemOpen
  \bibfield  {author} {\bibinfo {author} {\bibfnamefont {M.}~\bibnamefont
  {Punturo}}, \bibinfo {author} {\bibfnamefont {M.}~\bibnamefont {Abernathy}},
  \bibinfo {author} {\bibfnamefont {F.}~\bibnamefont {Acernese}}, \bibinfo
  {author} {\bibfnamefont {B.}~\bibnamefont {Allen}}, \bibinfo {author}
  {\bibfnamefont {N.}~\bibnamefont {Andersson}},  \emph {et~al.},\ }\href
  {\doibase 10.1088/0264-9381/27/19/194002} {\bibfield  {journal} {\bibinfo
  {journal} {Class.Quant.Grav.}\ }\textbf {\bibinfo {volume} {27}},\ \bibinfo
  {pages} {194002} (\bibinfo {year} {2010}{\natexlab{a}})}\BibitemShut
  {NoStop}%
\bibitem [{\citenamefont {Punturo}\ \emph
  {et~al.}(2010{\natexlab{b}})\citenamefont {Punturo}, \citenamefont
  {Abernathy}, \citenamefont {Acernese}, \citenamefont {Allen}, \citenamefont
  {Andersson} \emph {et~al.}}]{Punturo:2010zza}%
  \BibitemOpen
  \bibfield  {author} {\bibinfo {author} {\bibfnamefont {M.}~\bibnamefont
  {Punturo}}, \bibinfo {author} {\bibfnamefont {M.}~\bibnamefont {Abernathy}},
  \bibinfo {author} {\bibfnamefont {F.}~\bibnamefont {Acernese}}, \bibinfo
  {author} {\bibfnamefont {B.}~\bibnamefont {Allen}}, \bibinfo {author}
  {\bibfnamefont {N.}~\bibnamefont {Andersson}},  \emph {et~al.},\ }\href
  {\doibase 10.1088/0264-9381/27/8/084007} {\bibfield  {journal} {\bibinfo
  {journal} {Class.Quant.Grav.}\ }\textbf {\bibinfo {volume} {27}},\ \bibinfo
  {pages} {084007} (\bibinfo {year} {2010}{\natexlab{b}})}\BibitemShut
  {NoStop}%
\bibitem [{\citenamefont {Maggiore}\ \emph {et~al.}(2020)\citenamefont
  {Maggiore} \emph {et~al.}}]{Maggiore:2019uih}%
  \BibitemOpen
  \bibfield  {author} {\bibinfo {author} {\bibfnamefont {M.}~\bibnamefont
  {Maggiore}} \emph {et~al.},\ }\href {\doibase 10.1088/1475-7516/2020/03/050}
  {\bibfield  {journal} {\bibinfo  {journal} {JCAP}\ }\textbf {\bibinfo
  {volume} {03}},\ \bibinfo {pages} {050} (\bibinfo {year} {2020})},\ \Eprint
  {http://arxiv.org/abs/1912.02622} {arXiv:1912.02622 [astro-ph.CO]}
  \BibitemShut {NoStop}%
\bibitem [{\citenamefont {Sathyaprakash}\ \emph {et~al.}(2011)\citenamefont
  {Sathyaprakash} \emph {et~al.}}]{Sathyaprakash:2011bh}%
  \BibitemOpen
  \bibfield  {author} {\bibinfo {author} {\bibfnamefont {B.}~\bibnamefont
  {Sathyaprakash}} \emph {et~al.},\ }in\ \href@noop {} {\emph {\bibinfo
  {booktitle} {{46th Rencontres de Moriond on Gravitational Waves and
  Experimental Gravity}}}}\ (\bibinfo {year} {2011})\ pp.\ \bibinfo {pages}
  {127--136},\ \Eprint {http://arxiv.org/abs/1108.1423} {arXiv:1108.1423
  [gr-qc]} \BibitemShut {NoStop}%
\bibitem [{\citenamefont {Sathyaprakash}\ \emph {et~al.}(2012)\citenamefont
  {Sathyaprakash} \emph {et~al.}}]{Sathyaprakash:2012jk}%
  \BibitemOpen
  \bibfield  {author} {\bibinfo {author} {\bibfnamefont {B.}~\bibnamefont
  {Sathyaprakash}} \emph {et~al.},\ }\href {\doibase
  10.1088/0264-9381/29/12/124013} {\bibfield  {journal} {\bibinfo  {journal}
  {Class. Quant. Grav.}\ }\textbf {\bibinfo {volume} {29}},\ \bibinfo {pages}
  {124013} (\bibinfo {year} {2012})},\ \bibinfo {note} {[Erratum:
  Class.Quant.Grav. 30, 079501 (2013)]},\ \Eprint
  {http://arxiv.org/abs/1206.0331} {arXiv:1206.0331 [gr-qc]} \BibitemShut
  {NoStop}%
\bibitem [{\citenamefont {Amann}\ \emph {et~al.}(2020)\citenamefont {Amann}
  \emph {et~al.}}]{Amann:2020jgo}%
  \BibitemOpen
  \bibfield  {author} {\bibinfo {author} {\bibfnamefont {F.}~\bibnamefont
  {Amann}} \emph {et~al.},\ }\href {\doibase 10.1063/5.0018414} {\bibfield
  {journal} {\bibinfo  {journal} {Rev. Sci. Instrum.}\ }\textbf {\bibinfo
  {volume} {91}},\ \bibinfo {pages} {9} (\bibinfo {year} {2020})},\ \Eprint
  {http://arxiv.org/abs/2003.03434} {arXiv:2003.03434 [physics.ins-det]}
  \BibitemShut {NoStop}%
\bibitem [{\citenamefont {Radice}\ \emph {et~al.}(2017)\citenamefont {Radice},
  \citenamefont {Bernuzzi}, \citenamefont {Del~Pozzo}, \citenamefont
  {Roberts},\ and\ \citenamefont {Ott}}]{Radice:2016rys}%
  \BibitemOpen
  \bibfield  {author} {\bibinfo {author} {\bibfnamefont {D.}~\bibnamefont
  {Radice}}, \bibinfo {author} {\bibfnamefont {S.}~\bibnamefont {Bernuzzi}},
  \bibinfo {author} {\bibfnamefont {W.}~\bibnamefont {Del~Pozzo}}, \bibinfo
  {author} {\bibfnamefont {L.~F.}\ \bibnamefont {Roberts}}, \ and\ \bibinfo
  {author} {\bibfnamefont {C.~D.}\ \bibnamefont {Ott}},\ }\href {\doibase
  10.3847/2041-8213/aa775f} {\bibfield  {journal} {\bibinfo  {journal}
  {Astrophys. J.}\ }\textbf {\bibinfo {volume} {842}},\ \bibinfo {pages} {L10}
  (\bibinfo {year} {2017})},\ \Eprint {http://arxiv.org/abs/1612.06429}
  {arXiv:1612.06429 [astro-ph.HE]} \BibitemShut {NoStop}%
\bibitem [{\citenamefont {Buchner}(2021)}]{Buchner2021}%
  \BibitemOpen
  \bibfield  {author} {\bibinfo {author} {\bibfnamefont {J.}~\bibnamefont
  {Buchner}},\ }\href {\doibase 10.21105/joss.03001} {\bibfield  {journal}
  {\bibinfo  {journal} {Journal of Open Source Software}\ }\textbf {\bibinfo
  {volume} {6}},\ \bibinfo {pages} {3001} (\bibinfo {year} {2021})}\BibitemShut
  {NoStop}%
\bibitem [{\citenamefont {Skilling}(2006)}]{Skilling:2006}%
  \BibitemOpen
  \bibfield  {author} {\bibinfo {author} {\bibfnamefont {J.}~\bibnamefont
  {Skilling}},\ }\href {\doibase 10.1214/06-BA127} {\bibfield  {journal}
  {\bibinfo  {journal} {Bayesian Anal.}\ }\textbf {\bibinfo {volume} {1}},\
  \bibinfo {pages} {833} (\bibinfo {year} {2006})}\BibitemShut {NoStop}%
\bibitem [{\citenamefont {Kass}\ and\ \citenamefont
  {Raftery}(1995)}]{Kass:1995}%
  \BibitemOpen
  \bibfield  {author} {\bibinfo {author} {\bibfnamefont {R.~E.}\ \bibnamefont
  {Kass}}\ and\ \bibinfo {author} {\bibfnamefont {A.~E.}\ \bibnamefont
  {Raftery}},\ }\href {\doibase 10.1080/01621459.1995.10476572} {\bibfield
  {journal} {\bibinfo  {journal} {Journal of the American Statistical
  Association}\ }\textbf {\bibinfo {volume} {90}},\ \bibinfo {pages} {773}
  (\bibinfo {year} {1995})},\ \Eprint
  {http://arxiv.org/abs/https://amstat.tandfonline.com/doi/pdf/10.1080/01621459.1995.10476572}
  {https://amstat.tandfonline.com/doi/pdf/10.1080/01621459.1995.10476572}
  \BibitemShut {NoStop}%
\bibitem [{\citenamefont {Callister}(2021)}]{Callister:2021gxf}%
  \BibitemOpen
  \bibfield  {author} {\bibinfo {author} {\bibfnamefont {T.}~\bibnamefont
  {Callister}},\ }\href@noop {} {\  (\bibinfo {year} {2021})},\ \Eprint
  {http://arxiv.org/abs/2104.09508} {arXiv:2104.09508 [gr-qc]} \BibitemShut
  {NoStop}%
\bibitem [{\citenamefont {Zappa}\ \emph {et~al.}(2018)\citenamefont {Zappa},
  \citenamefont {Bernuzzi}, \citenamefont {Radice}, \citenamefont {Perego},\
  and\ \citenamefont {Dietrich}}]{Zappa:2017xba}%
  \BibitemOpen
  \bibfield  {author} {\bibinfo {author} {\bibfnamefont {F.}~\bibnamefont
  {Zappa}}, \bibinfo {author} {\bibfnamefont {S.}~\bibnamefont {Bernuzzi}},
  \bibinfo {author} {\bibfnamefont {D.}~\bibnamefont {Radice}}, \bibinfo
  {author} {\bibfnamefont {A.}~\bibnamefont {Perego}}, \ and\ \bibinfo {author}
  {\bibfnamefont {T.}~\bibnamefont {Dietrich}},\ }\href {\doibase
  10.1103/PhysRevLett.120.111101} {\bibfield  {journal} {\bibinfo  {journal}
  {Phys. Rev. Lett.}\ }\textbf {\bibinfo {volume} {120}},\ \bibinfo {pages}
  {111101} (\bibinfo {year} {2018})},\ \Eprint
  {http://arxiv.org/abs/1712.04267} {arXiv:1712.04267 [gr-qc]} \BibitemShut
  {NoStop}%
\bibitem [{\citenamefont {Bernuzzi}\ \emph {et~al.}(2016)\citenamefont
  {Bernuzzi}, \citenamefont {Radice}, \citenamefont {Ott}, \citenamefont
  {Roberts}, \citenamefont {Moesta},\ and\ \citenamefont
  {Galeazzi}}]{Bernuzzi:2015opx}%
  \BibitemOpen
  \bibfield  {author} {\bibinfo {author} {\bibfnamefont {S.}~\bibnamefont
  {Bernuzzi}}, \bibinfo {author} {\bibfnamefont {D.}~\bibnamefont {Radice}},
  \bibinfo {author} {\bibfnamefont {C.~D.}\ \bibnamefont {Ott}}, \bibinfo
  {author} {\bibfnamefont {L.~F.}\ \bibnamefont {Roberts}}, \bibinfo {author}
  {\bibfnamefont {P.}~\bibnamefont {Moesta}}, \ and\ \bibinfo {author}
  {\bibfnamefont {F.}~\bibnamefont {Galeazzi}},\ }\href {\doibase
  10.1103/PhysRevD.94.024023} {\bibfield  {journal} {\bibinfo  {journal} {Phys.
  Rev.}\ }\textbf {\bibinfo {volume} {D94}},\ \bibinfo {pages} {024023}
  (\bibinfo {year} {2016})},\ \Eprint {http://arxiv.org/abs/1512.06397}
  {arXiv:1512.06397 [gr-qc]} \BibitemShut {NoStop}%
\bibitem [{\citenamefont {Prakash}\ \emph {et~al.}(2021)\citenamefont
  {Prakash}, \citenamefont {Radice}, \citenamefont {Logoteta}, \citenamefont
  {Perego}, \citenamefont {Nedora}, \citenamefont {Bombaci}, \citenamefont
  {Kashyap}, \citenamefont {Bernuzzi},\ and\ \citenamefont
  {Endrizzi}}]{Prakash:2021wpz}%
  \BibitemOpen
  \bibfield  {author} {\bibinfo {author} {\bibfnamefont {A.}~\bibnamefont
  {Prakash}}, \bibinfo {author} {\bibfnamefont {D.}~\bibnamefont {Radice}},
  \bibinfo {author} {\bibfnamefont {D.}~\bibnamefont {Logoteta}}, \bibinfo
  {author} {\bibfnamefont {A.}~\bibnamefont {Perego}}, \bibinfo {author}
  {\bibfnamefont {V.}~\bibnamefont {Nedora}}, \bibinfo {author} {\bibfnamefont
  {I.}~\bibnamefont {Bombaci}}, \bibinfo {author} {\bibfnamefont
  {R.}~\bibnamefont {Kashyap}}, \bibinfo {author} {\bibfnamefont
  {S.}~\bibnamefont {Bernuzzi}}, \ and\ \bibinfo {author} {\bibfnamefont
  {A.}~\bibnamefont {Endrizzi}},\ }\href {\doibase 10.1103/PhysRevD.104.083029}
  {\bibfield  {journal} {\bibinfo  {journal} {Phys. Rev. D}\ }\textbf {\bibinfo
  {volume} {104}},\ \bibinfo {pages} {083029} (\bibinfo {year} {2021})},\
  \Eprint {http://arxiv.org/abs/2106.07885} {arXiv:2106.07885 [astro-ph.HE]}
  \BibitemShut {NoStop}%
\bibitem [{\citenamefont {Perego}\ \emph {et~al.}(2019)\citenamefont {Perego},
  \citenamefont {Bernuzzi},\ and\ \citenamefont {Radice}}]{Perego:2019adq}%
  \BibitemOpen
  \bibfield  {author} {\bibinfo {author} {\bibfnamefont {A.}~\bibnamefont
  {Perego}}, \bibinfo {author} {\bibfnamefont {S.}~\bibnamefont {Bernuzzi}}, \
  and\ \bibinfo {author} {\bibfnamefont {D.}~\bibnamefont {Radice}},\ }\href
  {\doibase 10.1140/epja/i2019-12810-7} {\bibfield  {journal} {\bibinfo
  {journal} {Eur. Phys. J.}\ }\textbf {\bibinfo {volume} {A55}},\ \bibinfo
  {pages} {124} (\bibinfo {year} {2019})},\ \Eprint
  {http://arxiv.org/abs/1903.07898} {arXiv:1903.07898 [gr-qc]} \BibitemShut
  {NoStop}%
\bibitem [{\citenamefont {Bernuzzi}\ \emph {et~al.}(2020)\citenamefont
  {Bernuzzi} \emph {et~al.}}]{Bernuzzi:2020txg}%
  \BibitemOpen
  \bibfield  {author} {\bibinfo {author} {\bibfnamefont {S.}~\bibnamefont
  {Bernuzzi}} \emph {et~al.},\ }\href {\doibase 10.1093/mnras/staa1860}
  {\bibfield  {journal} {\bibinfo  {journal} {Mon. Not. Roy. Astron. Soc.}\ }
  (\bibinfo {year} {2020}),\ 10.1093/mnras/staa1860},\ \Eprint
  {http://arxiv.org/abs/2003.06015} {arXiv:2003.06015 [astro-ph.HE]}
  \BibitemShut {NoStop}%
\bibitem [{\citenamefont {Radice}\ and\ \citenamefont
  {Rezzolla}(2012)}]{Radice:2012cu}%
  \BibitemOpen
  \bibfield  {author} {\bibinfo {author} {\bibfnamefont {D.}~\bibnamefont
  {Radice}}\ and\ \bibinfo {author} {\bibfnamefont {L.}~\bibnamefont
  {Rezzolla}},\ }\href {\doibase 10.1051/0004-6361/201219735} {\bibfield
  {journal} {\bibinfo  {journal} {Astron. Astrophys.}\ }\textbf {\bibinfo
  {volume} {547}},\ \bibinfo {pages} {A26} (\bibinfo {year} {2012})},\ \Eprint
  {http://arxiv.org/abs/1206.6502} {arXiv:1206.6502 [astro-ph.IM]} \BibitemShut
  {NoStop}%
\bibitem [{\citenamefont {Stergioulas}\ \emph {et~al.}(2004)\citenamefont
  {Stergioulas}, \citenamefont {Apostolatos},\ and\ \citenamefont
  {Font}}]{Stergioulas:2003ep}%
  \BibitemOpen
  \bibfield  {author} {\bibinfo {author} {\bibfnamefont {N.}~\bibnamefont
  {Stergioulas}}, \bibinfo {author} {\bibfnamefont {T.~A.}\ \bibnamefont
  {Apostolatos}}, \ and\ \bibinfo {author} {\bibfnamefont {J.~A.}\ \bibnamefont
  {Font}},\ }\href@noop {} {\bibfield  {journal} {\bibinfo  {journal} {Mon.
  Not. Roy. Astron. Soc.}\ }\textbf {\bibinfo {volume} {352}},\ \bibinfo
  {pages} {1089} (\bibinfo {year} {2004})},\ \Eprint
  {http://arxiv.org/abs/astro-ph/0312648} {astro-ph/0312648} \BibitemShut
  {NoStop}%
\bibitem [{\citenamefont {Bernuzzi}\ and\ \citenamefont
  {Nagar}(2008)}]{Bernuzzi:2008fu}%
  \BibitemOpen
  \bibfield  {author} {\bibinfo {author} {\bibfnamefont {S.}~\bibnamefont
  {Bernuzzi}}\ and\ \bibinfo {author} {\bibfnamefont {A.}~\bibnamefont
  {Nagar}},\ }\href {\doibase 10.1103/PhysRevD.78.024024} {\bibfield  {journal}
  {\bibinfo  {journal} {Phys. Rev.}\ }\textbf {\bibinfo {volume} {D78}},\
  \bibinfo {pages} {024024} (\bibinfo {year} {2008})},\ \Eprint
  {http://arxiv.org/abs/0803.3804} {arXiv:0803.3804 [gr-qc]} \BibitemShut
  {NoStop}%
\bibitem [{\citenamefont {Bernuzzi}(2009)}]{Bernuzzi:2009sak}%
  \BibitemOpen
  \bibfield  {author} {\bibinfo {author} {\bibfnamefont {S.}~\bibnamefont
  {Bernuzzi}},\ }\emph {\bibinfo {title} {{Numerical simulations of
  relativistic star oscillations: Gravitational waveforms from perturbative and
  3-dimensional codes}}},\ \href
  {http://inspirehep.net/record/1635979/files/fulltext.pdf} {Ph.D. thesis},\
  \bibinfo  {school} {Parma U.} (\bibinfo {year} {2009})\BibitemShut {NoStop}%
\bibitem [{\citenamefont {Radice}\ \emph {et~al.}(2010)\citenamefont {Radice},
  \citenamefont {Rezzolla},\ and\ \citenamefont {Kellermann}}]{Radice:2010rw}%
  \BibitemOpen
  \bibfield  {author} {\bibinfo {author} {\bibfnamefont {D.}~\bibnamefont
  {Radice}}, \bibinfo {author} {\bibfnamefont {L.}~\bibnamefont {Rezzolla}}, \
  and\ \bibinfo {author} {\bibfnamefont {T.}~\bibnamefont {Kellermann}},\
  }\href {\doibase 10.1088/0264-9381/27/23/235015} {\bibfield  {journal}
  {\bibinfo  {journal} {Class. Quant. Grav.}\ }\textbf {\bibinfo {volume}
  {27}},\ \bibinfo {pages} {235015} (\bibinfo {year} {2010})},\ \Eprint
  {http://arxiv.org/abs/1007.2809} {arXiv:1007.2809 [gr-qc]} \BibitemShut
  {NoStop}%
\bibitem [{\citenamefont {Stergioulas}\ \emph {et~al.}(2011)\citenamefont
  {Stergioulas}, \citenamefont {Bauswein}, \citenamefont {Zagkouris},\ and\
  \citenamefont {Janka}}]{Stergioulas:2011gd}%
  \BibitemOpen
  \bibfield  {author} {\bibinfo {author} {\bibfnamefont {N.}~\bibnamefont
  {Stergioulas}}, \bibinfo {author} {\bibfnamefont {A.}~\bibnamefont
  {Bauswein}}, \bibinfo {author} {\bibfnamefont {K.}~\bibnamefont {Zagkouris}},
  \ and\ \bibinfo {author} {\bibfnamefont {H.-T.}\ \bibnamefont {Janka}},\
  }\href {\doibase 10.1111/j.1365-2966.2011.19493.x} {\bibfield  {journal}
  {\bibinfo  {journal} {Mon.Not.Roy.Astron.Soc.}\ }\textbf {\bibinfo {volume}
  {418}},\ \bibinfo {pages} {427} (\bibinfo {year} {2011})},\ \Eprint
  {http://arxiv.org/abs/1105.0368} {arXiv:1105.0368 [gr-qc]} \BibitemShut
  {NoStop}%
\bibitem [{\citenamefont {Banik}\ \emph {et~al.}(2014)\citenamefont {Banik},
  \citenamefont {Hempel},\ and\ \citenamefont {Bandyopadhyay}}]{Banik:2014qja}%
  \BibitemOpen
  \bibfield  {author} {\bibinfo {author} {\bibfnamefont {S.}~\bibnamefont
  {Banik}}, \bibinfo {author} {\bibfnamefont {M.}~\bibnamefont {Hempel}}, \
  and\ \bibinfo {author} {\bibfnamefont {D.}~\bibnamefont {Bandyopadhyay}},\
  }\href {\doibase 10.1088/0067-0049/214/2/22} {\bibfield  {journal} {\bibinfo
  {journal} {Astrophys. J. Suppl.}\ }\textbf {\bibinfo {volume} {214}},\
  \bibinfo {pages} {22} (\bibinfo {year} {2014})},\ \Eprint
  {http://arxiv.org/abs/1404.6173} {arXiv:1404.6173 [astro-ph.HE]} \BibitemShut
  {NoStop}%
\bibitem [{\citenamefont {Fujimoto}\ \emph {et~al.}(2022)\citenamefont
  {Fujimoto}, \citenamefont {Fukushima}, \citenamefont {Hotokezaka},\ and\
  \citenamefont {Kyutoku}}]{Fujimoto:2022xhv}%
  \BibitemOpen
  \bibfield  {author} {\bibinfo {author} {\bibfnamefont {Y.}~\bibnamefont
  {Fujimoto}}, \bibinfo {author} {\bibfnamefont {K.}~\bibnamefont {Fukushima}},
  \bibinfo {author} {\bibfnamefont {K.}~\bibnamefont {Hotokezaka}}, \ and\
  \bibinfo {author} {\bibfnamefont {K.}~\bibnamefont {Kyutoku}},\ }\href@noop
  {} {\  (\bibinfo {year} {2022})},\ \Eprint {http://arxiv.org/abs/2205.03882}
  {arXiv:2205.03882 [astro-ph.HE]} \BibitemShut {NoStop}%
\bibitem [{\citenamefont {Abbott}\ \emph {et~al.}(2020)\citenamefont {Abbott}
  \emph {et~al.}}]{Abbott:2020uma}%
  \BibitemOpen
  \bibfield  {author} {\bibinfo {author} {\bibfnamefont {B.}~\bibnamefont
  {Abbott}} \emph {et~al.} (\bibinfo {collaboration} {LIGO Scientific,
  Virgo}),\ }\href {\doibase 10.3847/2041-8213/ab75f5} {\bibfield  {journal}
  {\bibinfo  {journal} {Astrophys. J. Lett.}\ }\textbf {\bibinfo {volume}
  {892}},\ \bibinfo {pages} {L3} (\bibinfo {year} {2020})},\ \Eprint
  {http://arxiv.org/abs/2001.01761} {arXiv:2001.01761 [astro-ph.HE]}
  \BibitemShut {NoStop}%
\bibitem [{\citenamefont {Abbott}\ \emph {et~al.}(2021)\citenamefont {Abbott}
  \emph {et~al.}}]{LIGOScientific:2021psn}%
  \BibitemOpen
  \bibfield  {author} {\bibinfo {author} {\bibfnamefont {R.}~\bibnamefont
  {Abbott}} \emph {et~al.} (\bibinfo {collaboration} {LIGO Scientific, VIRGO,
  KAGRA}),\ }\href@noop {} {\  (\bibinfo {year} {2021})},\ \Eprint
  {http://arxiv.org/abs/2111.03634} {arXiv:2111.03634 [astro-ph.HE]}
  \BibitemShut {NoStop}%
\bibitem [{\citenamefont {Sekiguchi}\ \emph {et~al.}(2011)\citenamefont
  {Sekiguchi}, \citenamefont {Kiuchi}, \citenamefont {Kyutoku},\ and\
  \citenamefont {Shibata}}]{Sekiguchi:2011mc}%
  \BibitemOpen
  \bibfield  {author} {\bibinfo {author} {\bibfnamefont {Y.}~\bibnamefont
  {Sekiguchi}}, \bibinfo {author} {\bibfnamefont {K.}~\bibnamefont {Kiuchi}},
  \bibinfo {author} {\bibfnamefont {K.}~\bibnamefont {Kyutoku}}, \ and\
  \bibinfo {author} {\bibfnamefont {M.}~\bibnamefont {Shibata}},\ }\href
  {\doibase 10.1103/PhysRevLett.107.211101} {\bibfield  {journal} {\bibinfo
  {journal} {Phys.Rev.Lett.}\ }\textbf {\bibinfo {volume} {107}},\ \bibinfo
  {pages} {211101} (\bibinfo {year} {2011})},\ \Eprint
  {http://arxiv.org/abs/1110.4442} {arXiv:1110.4442 [astro-ph.HE]} \BibitemShut
  {NoStop}%
\bibitem [{\citenamefont {Bauswein}\ \emph {et~al.}(2019)\citenamefont
  {Bauswein}, \citenamefont {Bastian}, \citenamefont {Blaschke}, \citenamefont
  {Chatziioannou}, \citenamefont {Clark}, \citenamefont {Fischer},\ and\
  \citenamefont {Oertel}}]{Bauswein:2018bma}%
  \BibitemOpen
  \bibfield  {author} {\bibinfo {author} {\bibfnamefont {A.}~\bibnamefont
  {Bauswein}}, \bibinfo {author} {\bibfnamefont {N.-U.~F.}\ \bibnamefont
  {Bastian}}, \bibinfo {author} {\bibfnamefont {D.~B.}\ \bibnamefont
  {Blaschke}}, \bibinfo {author} {\bibfnamefont {K.}~\bibnamefont
  {Chatziioannou}}, \bibinfo {author} {\bibfnamefont {J.~A.}\ \bibnamefont
  {Clark}}, \bibinfo {author} {\bibfnamefont {T.}~\bibnamefont {Fischer}}, \
  and\ \bibinfo {author} {\bibfnamefont {M.}~\bibnamefont {Oertel}},\ }\href
  {\doibase 10.1103/PhysRevLett.122.061102} {\bibfield  {journal} {\bibinfo
  {journal} {Phys. Rev. Lett.}\ }\textbf {\bibinfo {volume} {122}},\ \bibinfo
  {pages} {061102} (\bibinfo {year} {2019})},\ \Eprint
  {http://arxiv.org/abs/1809.01116} {arXiv:1809.01116 [astro-ph.HE]}
  \BibitemShut {NoStop}%
\end{thebibliography}
\end{document}